\documentclass[12pt,preprint]{aastex}

\shorttitle{Steady Nature of Line-Driven Disk Winds}

\shortauthors{Pereyra et al.}

\begin{document}

\title{Date: \today \break
       On the Steady Nature of Line-Driven Disk Winds}

\author{Nicolas A. Pereyra\altaffilmark{1}}
\affil{University of Pittsburgh, Department of Physics and Astronomy,
       Pittsburgh, PA 15260}
\email{pereyra@bruno.phyast.pitt.edu}

\author{Stanley P. Owocki}
\affil{Bartol Research Institute, University of Delaware,
       Newark, DE 19716}
\email{owocki@bartol.udel.edu}

\author{D. John Hillier 
        and David A. Turnshek}
\affil{University of Pittsburgh, Department of Physics and Astronomy,
       Pittsburgh, PA 15260}
\email{jdh@galah.phyast.pitt.edu,
       turnshek@quasar.phyast.pitt.edu}

\begin{abstract}
We perform an analytic investigation of the stability of line-driven
disk winds,
independent of hydrodynamic simulations.
Our motive is to determine whether or not line-driven disk winds can
account for the wide/broad UV resonance absorption lines seen in
cataclysmic variables (CVs) and quasi-stellar objects (QSOs).
In both CVs and QSOs observations generally indicate that the
absorption arising in the outflowing winds has a steady velocity
structure on time scales exceeding years
(for CVs)
and decades
(for QSOs).
However,
published results from hydrodynamic simulations of line-driven disk
winds are mixed,
with some researchers claiming that the models are inherently unsteady,
while other models produce steady winds.
The analytic investigation presented here shows that if the accretion
disk is steady,
then the line-driven disk wind emanating from it can also be steady.
In particular,
we show that a gravitational force initially increasing along the wind
streamline,
which is characteristic of disk winds,
does not imply an unsteady wind.
The steady nature of line-driven disk winds is consistent with the 1D
streamline disk-wind models of Murray and collaborators and the 2.5D
time-dependent models of Pereyra and collaborators.
This paper emphasizes the underlying physics behind the steady nature
of line-driven disk winds using mathematically simple models that mimic
the disk environment.
\end{abstract}


\section{Introduction}

Accretion disks are commonly believed to be present in both
cataclysmic variables
(CVs)
and quasi-stellar objects and active galactic nuclei (QSOs/AGN).
In both types of object blue-shifted absorption troughs in UV resonance
lines are sometimes present. 
Therefore,
a popular scenario put forth to explain them involves outflowing winds
emanating from an accretion disk,
with the winds giving rise to absorption troughs in the objects'
spectra when they are viewed along preferential sight-lines.
However,
while this scenario is well motivated,
the driving mechanism for the winds is still debatable.

In CVs,
the P-Cygni type line profiles are observed only at inclination angles
close to the pole
(i.e., at $\lesssim 65^\circ$,
 where $0^\circ$ corresponds to the disk rotation axis)
when there are high inferred mass accretion rates
\citep[e.g.,][]{war95}.
Given these requirements,
the obvious source of the wind material is the disk itself.
The terminal velocities of the blue-shifted absorption troughs lie in
the range
$3,000 - 5,000 \; {\rm km \; s}^{-1}$,
which is comparable to escape velocities in the inner disk regions
surrounding the white dwarf.

In QSOs,
broad absorption lines
(BALs)
from an outflowing wind are observed in
$>10\%$
of QSOs
\citep[e.g.,][]{hf03}.
Outflow velocities up to
$20,000 - 30,000 \; {\rm km \; s}^{-1}$
are common.
However,
it is not clear to what degree the detection of BALs is an
orientation-angle effect
\citep[e.g.,][]{elv00},
as is the case for CVs,
or if instead BAL QSOs are a distinct QSO type with some special
intrinsic or evolutionary properties
\citep[e.g.,][]{bor02}.
Nevertheless,
if accretion disks power QSOs/AGN,
some viewing angle effects are likely to be present.
\citet{tur84} cited evidence for and speculated that the outflowing BAL
gas may result from material being radially driven off an inner
rotating disk,
however the 1D model of
\citet{mur95}
was the first serious attempt to explain QSO/AGN observations using an
accretion disk wind.

Another property that CVs and QSOs have in common is the existence of 
persistent velocity structure in their absorption troughs
(when present)
over significantly long time scales.
This steady velocity structure,
with changes
$<10$ ${\rm km \; s}^{-1}$,
persists at least up to years and decades,
for CVs and QSOs,
respectively.
However,
changes in the depths of the absorption have been observed in both
classes of objects.
For some relevant observations showing the behaviors see
\citet{fron01}
and
\citet{har02}
for results on CVs and
\citet{fol87}
and
\citet{bar92} for results on QSOs.
The behavior of the absorption in Q1303+308
\citep[][]{fol87}
is possibly very illustrative,
with more recent data showing that the absorption depth has steadily
increased over 20 years while the underlying velocity structure
persists (Foltz, private communication).

To explain CVs
\citet{per97a}
developed 2.5D dynamical models of line-driven disk winds
\citep[see also][]{per97a,per97b,per00,per03}.
They concluded that steady wind solutions do exist and that wind
terminal velocities are approximately independent of disk luminosity,
similar to line-driven winds in early-type stars.
They found that the mass-loss rate increased with disk luminosity and
that rotational forces were important,
causing velocity streamlines to collide and reduce speed,
giving rise to an enhanced density region where the strongest
absorption occurs.
Although the approximations of single scattering and constant
ionization were made,
these models are generally consistent with observations and they show a
strong dependence on inclination angle.

\citet{pro98}
also developed 2.5D models for CVs
\citep[see also][]{pro99,pro02}.
Some of their results,
for example the magnitude of the wind velocities,
were similar to those obtained by \cite{per97a}.
However,
a significant difference between the two groups was that
\citet{pro98,pro99}
found unsteady flows characterized by large amplitude fluctuations in
velocity and density on very short time scales.

For QSOs/AGN the 1D models of
\citet{mur95} indicate that,
with an appropriate x-ray shielding mechanism,
a suitable accretion disk wind can be driven off by line radiation
pressure.
Unlike the case for CVs,
the wind streamlines are approximately parallel to the disk at high
velocities
(i.e., close to inclination angles near $90^\circ$).
The models of
\citet{mur95}
are able to account for the observed outflow velocities seen in BAL
QSOs,
the presence of detached absorption troughs seen in some BAL QSOs,
and the approximate fraction of QSOs observed to have BALs.
But the 2.5D QSO/AGN line-driven wind models of
\citet{pro00}
challenge this result.
Similar to their earlier CV disk-wind models,
\citet{pro00}
and
\cite{pro02}
report finding intrinsically unsteady winds. 

Proga and collaborators have suggested that since their unsteady
line-driven disk-wind models are inconsistent with observational
results,
``that a factor other than line-driving is much more likely to be
  decisive in powering these outflows''
\citep[][]{har02}
\footnote{
  In line-driven disk wind models the mass loss rate is expected to
  increase/decrease with increasing/decreasing
  disk luminosity.
  The observations presented by \citet{har02} indicate that observables
  like the CIV$\lambda1549$ absorption equivalent width do not scale
  with disk luminosity.
  However,
  this argument should not be used to invalidate line-driven disk wind
  models since the CIV$\lambda1549$ absorption equivalent width may not
  be a direct measure of mass loss rate,
  e.g., due to ionization changes.
}
and that
``indeed radiation pressure alone does not suffice to
  drive the observed hypersonic flow''
\citep[][]{pro03}.
However,
the line-driven disk-wind models of Pereyra and collaborators
\citep[see][]{hil02}
continue to find steady wind solutions,
similar to the earlier CV models of
\citet{per97a}
and
\citet{per00}.

Thus,
an impasse of sorts has developed with regard to the applicability of
line-driven accretion disk wind models to CVs and QSOs/AGN.
Clearly the observations indicate that steady winds do exist in these
objects.
However,
while one group believes that steady line-driven winds can be achieved,
the other group has advocated either abandoning this approach or
adopting one that incorporates additional physics
(e.g., magnetic fields)
into the problem.
Therefore, 
the objective of this work is to develop a series of semi-analytical
models,
independent of previous 2.5D dynamical efforts,
to test for the existence/nonexistence of steady disk winds.
{\it We find that steady winds can exist}
and here we emphasize the physics behind these solutions by
employing mathematically simple models designed to mimic a disk
environment.
In a subsequent paper we will present detailed numerical calculations.

In \S\ref{sec_general_equations} we present the notation and general
equations used in this paper.
We define the nozzle function in
\S\ref{sec_nozzle}
and discuss its relationship to steady wind solutions.
In \S4 we analyze the FSH02 model
\citep{fel02}.
This analysis clearly demonstrates that an increase in gravity along
a streamline,
which is characteristic of disk winds,
does not necessarily imply an unsteady wind.
In \S5 we present the standard simple models which demonstrate steady
winds can exist.
The summary and conclusions are presented in \S6.

\section{General Equations}
\label{sec_general_equations}

\subsection{Hydrodynamic Equations}
\label{sec_hydrodynamic_equations}

Up to now our studies have indicated that for typical CV and QSO
parameters temperature gradient terms do not produce significant
changes in overall dynamical results. Thus, throughout this paper
we assume that the wind is isothermal.

We use the Gayley ($\bar{Q}$,$\alpha$) line force parameterization
\citep{gay95} for this study.
This parameterization is used in place of the CAK
$k$
parameter
\citep{cas75}.
The
$\bar{Q}$
parameter has a direct physical interpretation
\citep{gay95}.
The CAK
$k$
and 
the Gayley
$\bar{Q}$
parameters are related through

\begin{equation}
\label{equ_gayley}
  k
=
  {1 \over 1 - \alpha}
  \left(v_{th} \over c \right)^\alpha
  \bar{Q}^{1-\alpha}
\;\;\;\; ,
\end{equation}

\noindent
where
$\alpha$
is the CAK line force parameter
\citep{cas75},
$v_{th}$
is the ion thermal velocity,
and
$c$
is the speed of light.

The time-dependent hydrodynamic equations for a 1D system are
\hfill \break
(1) the mass conservation equation,

\begin{equation}
\label{equ_time_mass}
  A {\partial \rho \over \partial t}
+ {\partial \over \partial z} (\rho V A)
= 0 
\;\;\;\; ,
\end{equation}

\noindent
(2) the momentum equation,

\begin{equation}
\label{equ_time_momentum}
  \rho {\partial V \over \partial t}
+ \rho V {\partial V \over \partial z}
=
- \rho B
+ \rho {\kappa_e \over c} {\cal F} {\bar{Q}^{1-\alpha} \over 1-\alpha}
  \left( {1 \over \rho \kappa_e c}
         {\partial V \over \partial z} \right)^\alpha
- {\partial P \over \partial z}
\;\;\;\; ,
\end{equation}

\noindent
and (3) the equation of state,

\begin{equation}
\label{equ_time_state}
  P
=
  \rho \, b^2
\;\;\;\; .
\end{equation}

\noindent
Here
$z$
is the independent spatial coordinate which corresponds to
the height above the disk
(or the distance from the center of the star for stellar wind models),
$t$
is the time,
$V$
is the velocity,
$A$
is the area which depends on
$z$,
$\rho$
is the density,
$B$
represents the body forces which in this case is the gravitational
plus continuum radiation force per mass,
$\kappa_e$
is the Thomson cross section per mass,
${\cal F}$ is the radiation flux,
$P$ is the pressure,
and
$b$
is the isothermal sound speed.

To simplify the momentum equation,
we define the line opacity weighted flux
$\gamma(z)$
as

\begin{equation}
\label{equ_gamma}
  \gamma(z)
\equiv
  {\kappa_e \over c} {\cal F}(z) {\bar{Q}^{1-\alpha} \over 1-\alpha}
    \left( {1 \over \kappa_e c} \right)^\alpha
\;\;\;\; .
\end{equation}

\noindent
The time-dependent momentum equation can then be written as

\begin{equation}
\label{equ_time_momentum2}
  \rho {\partial V \over \partial t}
+ \rho V {\partial V \over \partial z}
=
- \rho B
+ \rho \gamma
  \left( {1 \over \rho} {\partial V \over \partial z} \right)^\alpha
- {\partial P \over \partial z}
\;\;\;\; ,
\end{equation}

\noindent
where the dependence of
$\gamma$
on
$z$
is implicit.

From equation
(\ref{equ_time_mass})
the stationary mass conservation equation is then

\begin{equation}
\label{equ_stationary_mass}
  \rho \, V A
=
  \dot{M}
\;\;\;\; ,
\end{equation}

\noindent
where
$\dot{M}$
is the wind mass loss rate and the stationary momentum equation is

\begin{equation}
\label{equ_stationary_momentum}
  \rho V {d V \over d z}
=
- \rho B
+ \rho \gamma
  \left( {1 \over \rho} {d V \over d z} \right)^\alpha
- {d P \over d z}
\;\;\;\; .
\end{equation}

\subsection{Equation of Motion}
\label{sec_equation_of_motion}

Combining equations (\ref{equ_time_state}),
(\ref{equ_stationary_mass})
and
(\ref{equ_stationary_momentum}),
we find that the equation of motion is

\begin{equation}
\label{equ_motion_v}
  \left(1 - {b^2 \over V^2} \right) V \, {dV \over dz}
=
  - B
  + \gamma \left( {A \over \dot{M}} V \, {dV \over dz} \right)^\alpha
  + {b^2 \over A} {dA \over dz}
\;\;\;\; ,
\end{equation}

\noindent
and if we define

\begin{equation}
\label{equ_W}
  W
\equiv
  {V^2 \over 2}
\;\;\;\; ,
\end{equation}

\noindent
the equation of motion becomes

\begin{equation}
\label{equ_motion_W}
  \left(1 - {b^2 \over 2 W} \right) {dW \over dz}
=
  - B
  + \gamma \left( {A \over \dot{M}} {dW \over dz} \right)^\alpha
  + {b^2 \over A} {dA \over dz}
\;\;\;\; .
\end{equation}

\subsection{Scaling of Physical Parameters}

For each model,
we define a value of $r_0$,
$B_0$,
$A_0$,
and
$\gamma_0$
as the characteristic distance,
gravitational acceleration,
area,
and line-weighted opacity, respectively.
The normalized spatial coordinate,
$x$,
body force,
$g$,
and area,
$a$,
are

\begin{equation}
\label{equ_x}
  x
\equiv
  {z \over r_0}
\hskip 24pt
;
\hskip 24pt
  g
\equiv
  {B \over B_0}
\hskip 24pt
;
\hskip 24pt
  a
\equiv
  {A \over A_0}
\;\;\;\; .
\end{equation}

\noindent
The normalized line opacity weighted flux
$f$
is

\begin{equation}
\label{equ_f}
  f
\equiv
  {1 \over \alpha^\alpha (1-\alpha)^{1-\alpha}} {\gamma \over \gamma_0}
\;\;\;\; .
\end{equation}

\noindent
The characteristic value of the dependent variable
$W$
is

\begin{equation}
\label{equ_W0}
  W_0
\equiv
  B_0 r_0
\;\;\;\; .
\end{equation}

\noindent
Then the characteristic velocity is given by

\begin{equation}
  V_0
\equiv
  \sqrt{2 W_0}
\;\;\;\; .
\end{equation}

\noindent
Based on the CAK stellar wind formalism
\citep{cas75},
the characteristic wind mass loss rate is

\begin{equation}
\label{equ_mcak}
  \dot{M}_{CAK}
\equiv
  \alpha (1-\alpha)^{1 - \alpha \over \alpha}
  {A_0 \gamma_0^{1 \over \alpha} \over B_0^{1-\alpha \over \alpha}}
\;\;\;\; .
\end{equation}

The corresponding normalized variables are defined as

\begin{equation}
\label{equ_w}
  \omega
\equiv
  {W \over W_0}
\hskip 24pt
;
\hskip 24pt
  v
\equiv
  {V \over V_0}
\hskip 24pt
;
\hskip 24pt
  \dot{m}
\equiv
  {\dot{M} \over \dot{M}_{CAK}}
\;\;\;\; .
\end{equation}

\noindent
The normalized sound speed squared is 

\begin{equation}
\label{equ_s}
  s
\equiv
  {b^2 \over 2 W_0}
\;\;\;\; .
\end{equation}

Finally,
by introducing the scaling relations from equations 
(\ref{equ_x})-(\ref{equ_s}),
the equation of motion becomes

\begin{equation}
\label{equ_motion_w}
  \left(1-{s \over \omega} \right) {d \omega \over d x}
=
- g
+ f \left({a \over \dot{m}} {d\omega \over dx} \right)^\alpha
+ {2 s \over a} {da \over dx}
\;\;\;\; .
\end{equation}

\noindent
This is the form of the equation of motion discussed throughout
most of this paper.
We emphasize at this point that for typical stellar, CV, and QSO
parameters $s \ll 1$.
For this reason $s$ has little influence on the fundamental
characteristics of wind solutions,
and thus will simply taken to be zero in much of the remaining
analysis.

The question of the existence of a steady solution is then reduced to
determining whether a value of
$\dot{m}$
and a normalized function
$\omega(x)$
exists such that it satisfies the boundary conditions and
equation~(\ref{equ_motion_w}).
A steady solution for a 1D hydrodynamical model exists if and only if
equation~(\ref{equ_motion_w}) is integrable while simultaneously
satisfying the boundary conditions.
We note that once we show that steady solutions exist
(\S\ref{sec_nozzle}),
we then demonstrate their stability
(\S\ref{fsh02} and \S\ref{simple_models})
using numerical time-dependent hydrodynamical models based on the PPM
numerical scheme \citep{col84}.

\section{Nozzle Function and Critical Point}
\label{sec_nozzle}

Motivated by analogies between line-driven winds and a supersonic
nozzle \citep{abb80},
we have found that insights can be obtained by defining and considering
the nozzle function,
$n(x)$.
The relationship between the nozzle function, the critical point,
and the existence/nonexistence of a steady 1D wind solution is
elaborated in this section.
Since mathematical expressions associated with critical point
conditions have forms which readily allow for physical
interpretation,
we discuss the nozzle function and critical point conditions
simultaneously.

In \S\ref{sec_s0_a05} we initially analyze the simple case
where
$s=0$
and
$\alpha = 1/2$.
This case results in an explicit analytical expression for
$d\omega/dx$
from the equation of motion
[equation~(\ref{equ_motion_w})].
In \S\ref{sec_s0} we extend the results to include an arbitrary value
for
$\alpha$
($0<\alpha<1$).
In \S\ref{sec_sg0} we briefly discuss the case of a finite
sound speed ($s>0$).
We note that although gas pressure does not produce significant
corrections to wind mass loss rates and velocity laws for typical CV
and QSO parameters,
gas pressure effects do give rise to the necessity of a critical point
in order for the wind solution to be steady.
This result has been discussed in detail by
\citet{cas75}
for the case of line-driven stellar winds.
In \S\ref{sec_critical} we extend these arguments to the equation
of motion presented here
[equation(\ref{equ_motion_w})].
Finally,
in order too illustrate the application of the nozzle function to
line-driven winds,
in \S\ref{sec_cak} we apply it to the well-studied CAK stellar wind.

\subsection{$s = 0$ and $\alpha = 1/2$}
\label{sec_s0_a05}

For this case the equation of motion becomes

\begin{equation}
\label{equ_motion_s0_a05}
  {d \omega \over d x}
=
- g
+ f \left({a \over \dot{m}} {d\omega \over dx} \right)^{1/2}
\;\;\;\; , 
\end{equation}

\noindent
and the question of the existence of a steady solution is reduced
to the question of whether or not a value for
$d\omega / dx$
can always be obtained when integrating the equation of motion.

From equation~(\ref{equ_motion_s0_a05}),
one obtains

\begin{equation}
\label{equ_wp_s0_a05}
  {d\omega \over dx}
=
  {f^2 a \over 4 \dot{m}}
  \left[1 \pm \sqrt{1 - {4 \dot{m} g \over f^2 a}} \; \right]^2
\;\;\;\; .
\end{equation}

\noindent
The nozzle function is defined to be 

\begin{equation}
\label{equ_n_s0_a05}
  n(x)
\equiv
  {f^2(x) a(x) \over 4 g(x)}
\hbox{\hskip 48pt}
  \left[
=
   {(fa)^2 \over 4 (ga)}
  \right]
\hbox{\hskip 48pt}
\;\;\;\; .
\end{equation}

\noindent 
The expression for
$d\omega /dx$
can now be written as

\begin{equation}
\label{equ_wp_s0_a05_n}
  {d\omega \over dx}
=
  {n(x) g(x) \over \dot{m}}
  \left[1 \pm \sqrt{1 - { \dot{m} \over n(x)}} \; \right]^2
\;\;\;\; .
\end{equation}

\noindent
A steady solution must pass through a critical point,
$x_c$,
passing from a lower branch
[corresponding to the ``-'' sign in equation~(\ref{equ_wp_s0_a05_n})],
to a higher branch
[corresponding to the ``+'' sign in equation~(\ref{equ_wp_s0_a05_n})]
(see \S\ref{sec_critical} for details).
Therefore, 
it must hold that

\begin{equation}
  \dot{m}
=
  \min[n(x)]
\equiv
  n_c
\equiv
  n(x_c)
\;\;\;\; ,
\end{equation}

\noindent
where $n_c$ is the value of the nozzle function at the critical point.

Thus,
one finds that the wind mass loss rate is determined by the minimum
value of the nozzle function,
where the wind mass loss rate is the maximum possible value that
permits integration of the equation of motion
(i.e.,
 the maximum steady wind mass loss rate that the system can physically
 support).

The velocity law is obtained by integrating through the lower branch of
the equation of motion from the critical point down to the sonic point,
$x_s$,
and by integrating the equation of motion through the upper branch from
the critical point out to infinity.
The constants of integration are determined through the conditions of
continuity of the velocity law and the condition that
$w(x_s)=0$.

\begin{figure}
\epsscale{1.0}
\plotone{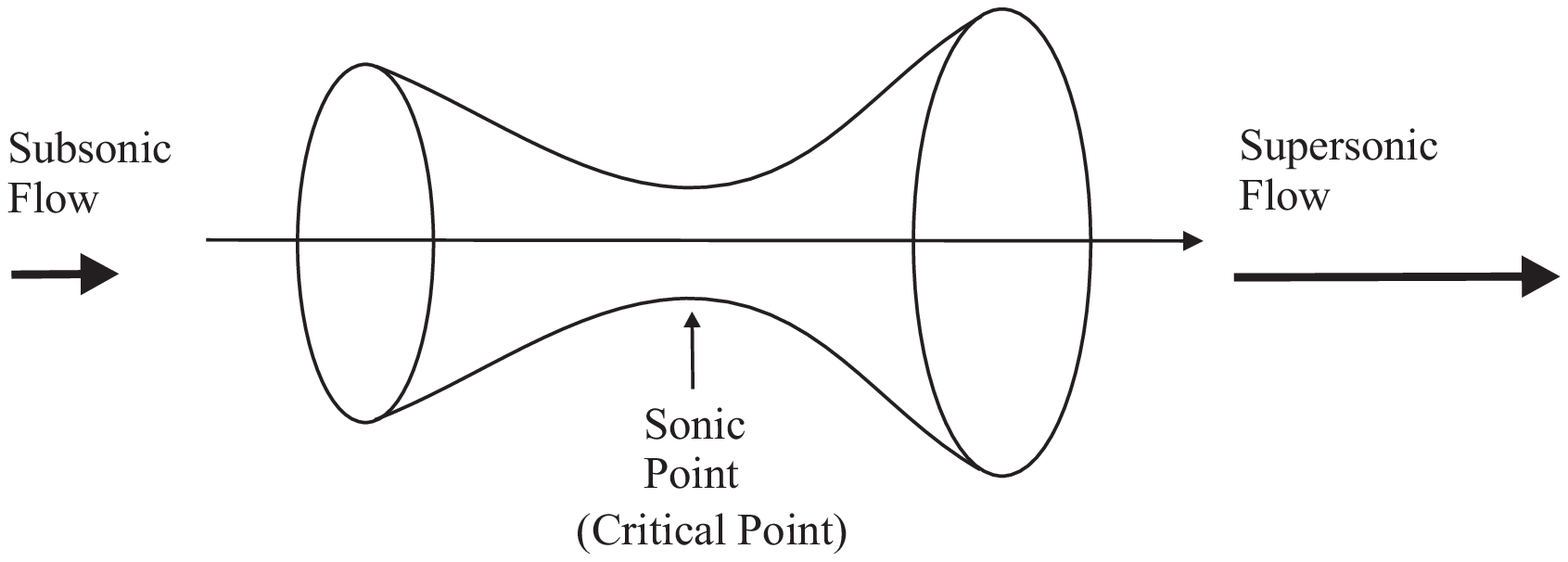}
\caption{
Supersonic Nozzle
}
\label{fig_nozzle}
\end{figure}

\citet{abb80}
discussed in detail the analogies between a stellar line-driven wind
and a supersonic nozzle
(a tube with a gas flow which starts subsonic and ends supersonic,
 see Figure~\ref{fig_nozzle}).
One similarity between a supersonic nozzle and the nozzle function of a
1D line-driven wind with negligible gas pressure is that a necessary
condition for a steady flow is that the cross sectional area must have
a minimum value within the nozzle
\cite[e.g.,][]{lan97}.
Also,
just as is the case with negligible gas pressure,
the nozzle function in a line-driven wind must also present a minimum
if a steady supersonic wind solution is to exist.

Another similarity,
also discussed by
\citet{abb80}
for stellar winds,
comes from the propagation velocity of density perturbations at the
critical point.
In a supersonic nozzle the sonic point,
which is at the minimum of the nozzle cross sectional area,
is the point where the flow velocity equals the propagation velocity of
density perturbations
(i.e., the sound speed).
In a line-driven wind the critical point is the point where the flow
velocity equals the backward velocity propagation of density
perturbations, referred to as radiative-acoustic waves or Abbott waves
(as opposed to sound waves).
Due to the dependence of the line radiation force on the velocity
gradient,
density perturbations in a line-driven wind will travel at velocities
different from sound speed.
The propagation velocity will be subsonic in the forward direction
(i.e., the direction of the line-driving force)
and supersonic in the backward direction.

\subsection{$s = 0$ and $0 < \alpha < 1$}
\label{sec_s0}

For this case the equation of motion becomes

\begin{equation}
\label{equ_motion_s0}
  {d \omega \over d x}
=
- g
+ f \left({a \over \dot{m}} {d\omega \over dx} \right)^{\alpha}
\;\;\;\; .
\end{equation}

\noindent
The question of the existence of a steady solution is again reduced to
the question of whether or not a value for
$d\omega / dx$
can always be obtained when integrating the equation of motion.

From an analysis similar to that presented by
\citet{cas75}
for line-driven stellar winds,
one finds that equation~(\ref{equ_motion_s0}) has two solution branches
which exist under the condition that

\begin{equation}
\label{equ_mdotn_s0}
  \dot{m}
\leq
  \alpha (1-\alpha)^{1 - \alpha \over \alpha}
    {f^{1 \over \alpha} a \over g^{1-\alpha \over \alpha}}
\;\;\;\; .
\end{equation}

\noindent
The two branches intersect at the critical point when both sides of
equation~(\ref{equ_mdotn_s0}) are equal.

This leads to the following definition of the nozzle function

\begin{equation}
\label{equ_n_s0}
  n(x)
\equiv
  \alpha (1-\alpha)^{1 - \alpha \over \alpha}
    {f^{1 \over \alpha} a \over g^{1-\alpha \over \alpha}} \;
\hbox{\hskip 48pt}
  \left[
=
  \alpha (1-\alpha)^{1 - \alpha \over \alpha}
    {(fa)^{1 \over \alpha} \over (ga)^{1-\alpha \over \alpha}} \;
  \right]
\hbox{\hskip 48pt}
\; 
\;\;\;\; .
\end{equation}

The condition for the existence of the two solutions branches can now
be written as

\begin{equation}
  \dot{m}
\leq
  n(x)
\;\;\;\; .
\end{equation}

\noindent
Again we require a critical point type solution
(see \S\ref{sec_critical})
and it follows that

\begin{equation}
  \dot{m}
=
  \min[n(x)]
\equiv
  n_c
\equiv
  n(x_c)
\;\;\;\; .
\end{equation}

\noindent
The wind mass loss rate and velocity law are determined just as in
the
$\alpha = 1/2$
case.

\subsection{CAK Model}
\label{sec_cak}

\begin{figure}
\epsscale{1.0}
\plotone{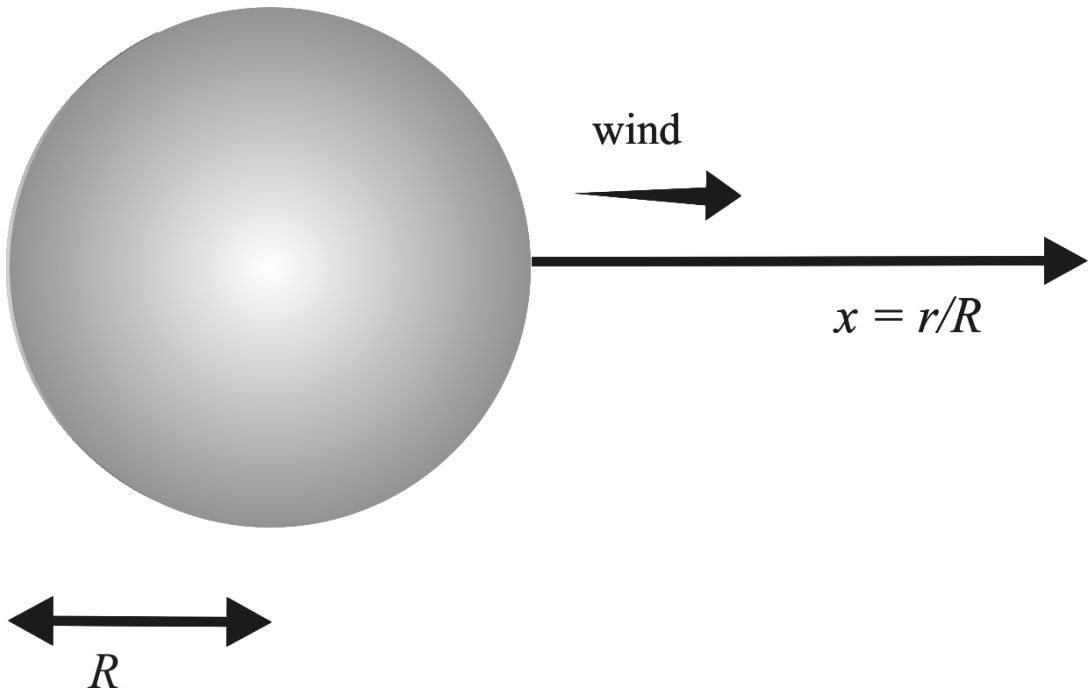}
\caption{
CAK Stellar Wind
}
\label{fig_cak}
\end{figure}

Here we illustrate the concepts and notations which we have
introduced by applying them to the well-known and well-studied CAK
stellar wind
\citep{cas75}.
For simplicity we consider the case of zero sound speed only
($s=0$).
For the CAK model
(see Figure~\ref{fig_cak})
we use the following characteristic scales

\begin{equation}
  r_0
=
  R
\hskip 24pt
;
\hskip 24pt
  B_0
=
  {G M \over R^2} (1-\Gamma)
\hskip 24pt
;
\hskip 24pt
  A_0
=
  4 \pi R^2
\;\;\;\; ;
\end{equation}

\noindent
and

\begin{equation}
  \gamma_0
=
  {\kappa_e \over c}
  {{\cal L} \over 4 \pi R^2}
  k
  \left(1 \over \kappa_e v_{th} \right)^\alpha
\;\;\;\; .
\end{equation}

\noindent
Here $R$ is the photospheric radius,
$G$
is the gravitational constant,
$M$
is the stellar mass,
${\cal L}$
is the stellar luminosity,
and
$\Gamma$
is the Eddington ratio given by

\begin{equation}
  \Gamma
=
  {\kappa_e {\cal L} \over 4 \pi G M c}
\;\;\;\; .
\end{equation}

\noindent
For the CAK model the independent spatial variable is the distance, 
$r$,
to the center of the star,
and thus
$x=r/R$.

With the corresponding substitutions,
the equation of motion
(for $s=0$)
becomes

\begin{equation}
  {d \omega \over dx}
=
- {1 \over x^2}
+ {1 \over \alpha^\alpha (1-\alpha)^{1-\alpha}} {1 \over x^2}
  \left({x^2 \over \dot{m}} {d\omega \over dx} \right)^\alpha
\;\;\;\; .
\end{equation}

\noindent
The expressions for the normalized variables for gravitational
acceleration,
line opacity weighted flux,
and area are,
respectively,

\begin{equation}
\label{equ_cak_g}
  g
=
  {1 \over x^2}
\;\;\;\; ;
\end{equation}

\begin{equation}
\label{equ_cak_f}
  f
=
  {1 \over \alpha^\alpha (1-\alpha)^{1-\alpha}} {1 \over x^2}
\;\;\;\; ;
\end{equation}

\noindent
and

\begin{equation}
\label{equ_cak_a}
  a
=
  x^2
\;\;\;\; ;
\end{equation}

\noindent
Substituting equations~(\ref{equ_cak_g})-(\ref{equ_cak_a}) into
equation~(\ref{equ_n_s0}),
we find

\begin{equation}
  n(x)
=
  1
\;\;\;\; .
\end{equation}

\noindent
This constancy of the nozzle function implies that all spatial points
become critical points
(i.e.,
      all points are a global minimum of the nozzle function).
It follows that the normalized wind mass loss rate is given by:

\begin{equation}
\label{equ_cak_mdot}
  \dot{m}
=
  1
\;\;\;\; .
\end{equation}

The equation of motion becomes

\begin{equation}
  \left(x^2 {d \omega \over dx}\right)
=
- 1
+ {1 \over \alpha^\alpha (1-\alpha)^{1-\alpha}}
  \left(x^2 {d\omega \over dx} \right)^\alpha
\;\;\;\; ,
\end{equation}

\noindent
which in turn implies

\begin{equation}
  x^2 {d\omega \over dx}
=
  {\alpha \over 1 - \alpha}
\;\;\;\; .
\end{equation}

\noindent
Integrating the last equation under the condition that
$\omega(1) = 0$
(zero velocity at photospheric height),
we obtain the normalized velocity law

\begin{equation}
\label{equ_cak_w}
  \omega
=
  {\alpha \over 1-\alpha} \left(1-{1\over x}\right)
\;\;\;\; .
\end{equation}

Expressing equations~(\ref{equ_cak_mdot}) and (\ref{equ_cak_w})
in terms of the stellar parameters we obtain

\begin{equation}
  \dot{M}
=
  { \alpha (1-\alpha)^{1-\alpha \over \alpha} \over
    (4\pi)^{1-\alpha \over \alpha} }
  { \left[ {\kappa_e \over c} {\cal L}
    {\bar{Q}^{1-\alpha} \over 1 - \alpha}
      \left(1 \over \kappa_e c \right)^\alpha
    \right]^{1 \over \alpha} \over
    [GM(1-\Gamma)]^{1-\alpha \over \alpha} }
\;\;\;\; ;
\end{equation}

\noindent
and

\begin{equation}
  V
=
  \left(\alpha \over 1-\alpha\right)^{ 1\over 2}
  \left(2GM(1-\Gamma) \over R\right)^{1\over 2}
  \left(1-{R\over r}\right)^{1\over 2}
\;\;\;\; .
\end{equation}

\noindent
These are the well-known expressions derived by
\citet{cas75}
for the wind mass loss rate
$\dot{M}$
and the wind velocity law
$V(r)$
in the asymptotic limit where
$V$
is much greater than the sound speed.

\section{FSH02 Model}
\label{fsh02}

Here we analyze the FSH02 model discussed by
\citet{fel02}.
The motivation for studying the FSH02 in this work is twofold.
First,
the analysis clearly shows that an increase in gravity along a
streamline,
which is characteristic of disk winds,
does not imply an unsteady wind.
Second,
for the case where $\alpha=1/2$,
an explicit analytical solution can be found which will serve as a
consistency check for the numerical codes.

The FSH02 model is defined through the following three
equations

\begin{equation}
  g(x)
=
  {x \over 1+x^2}
\;\;\;\; ;
\end{equation}

\begin{equation}
  a(x)
=
  1
\;\;\;\; ;
\end{equation}

\noindent
and

\begin{equation}
  f(x)
=
  1
\;\;\;\; ,
\end{equation}

\noindent
where for these disk-like models $x$ is the distance along the vertical
streamline normalized by the distance from the center of the disk.
Assuming $s=0$ and $\alpha=1/2$,
the equation of motion for the FSH02 model becomes

\begin{equation}
\label{equ_fs02_motion}
  {d \omega \over dx}
=
- {x \over 1+x^2}
+\left({1 \over \dot{m}} {d\omega \over dx} \right)^{1 \over 2}
\;\;\;\; .
\end{equation}

\noindent
From equation~(\ref{equ_n_s0_a05}),
the nozzle function for this model is given by

\begin{equation}
  n(x)
=
  {1 + x^2 \over 4 x}
\;\;\;\; .
\end{equation}

\noindent
It follows that:

\begin{equation}
  \min(n[x])
=
  {1 \over 2}
=
  n(1)
\hbox{\hskip 48pt}
(0 \leq x)
\;\;\;\; .
\end{equation}

\noindent
Therefore,

\begin{equation}
  x_c
=
  1 
\;\;\;\;\;\;\;\;
{\rm and}
\;\;\;\;\;\;\;\;
  \dot{m}
=
  {1 \over 2}
\;\;\;\; .
\end{equation}

\noindent
Substituting this value of
$\dot{m}$
into equation~(\ref{equ_fs02_motion}) and integrating,
with the additional condition that $\omega(0)=0$,
we find

\begin{equation}
  \omega(x)
=
  x - 1 + \sqrt{1+x^2} - {\rm arcsinh}(x) - {1\over 2}\ln(1+x^2)
\;\;\;\; .
\end{equation}

\begin{figure}
\plotone{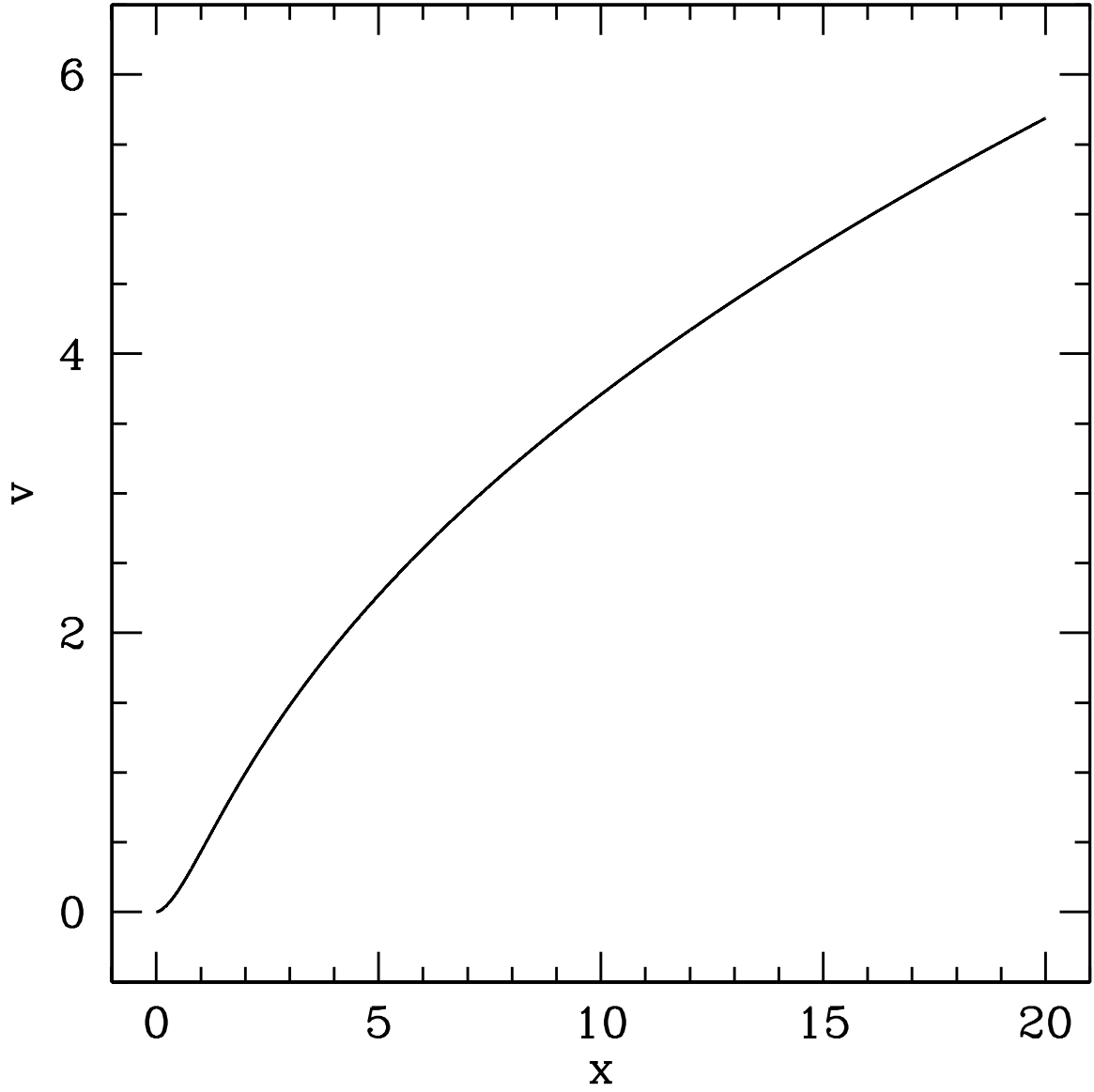}
\caption{
Analytic velocity law for the FSH02 model ($s=0$).
}
\label{fig_fs02_vth}
\end{figure}

\noindent
In Figure~\ref{fig_fs02_vth} we plot this velocity law in terms of
$v\;=\sqrt{\omega}\;$.

\begin{figure}
\epsscale{1.0}
\plotone{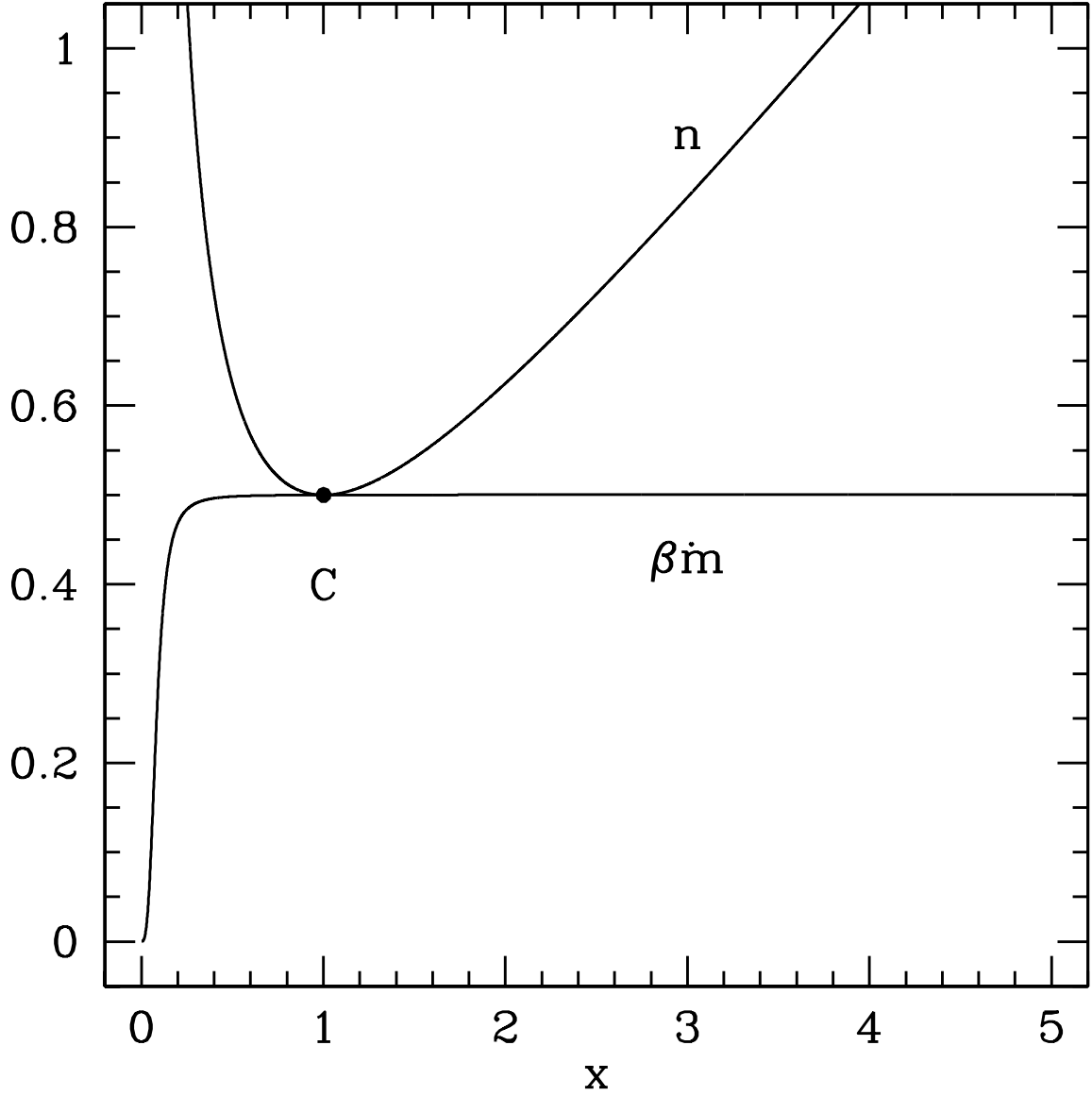}
\caption{
Nozzle function $n(x)$ and the product
$\beta(\omega[x]) \dot{m}$
for the FSH02 model ($s=10^{-4}$).
``C'' denotes the critical point.
}
\label{fig_fs02_b}
\end{figure}

Figure~\ref{fig_fs02_b} shows the nozzle function
$n(x)$
of the FSH02 model for
$s=10^{-4}$
along with the product
$\beta(\omega[x])
\dot{m}$.
These two graphs intersect at,
and only at,
the critical point.
Since
$\beta(\omega[x])$
is a monotonically increasing function,
the critical point must be (slightly) to the right of the nozzle
minimum.
Also,
since in the FSH02 model
$da/dx = 0$,
it follows that for this model in particular the nozzle function is
independent of
$s$
[see equation~(\ref{equ_n})].

\begin{figure}
\epsscale{1.0}
\plotone{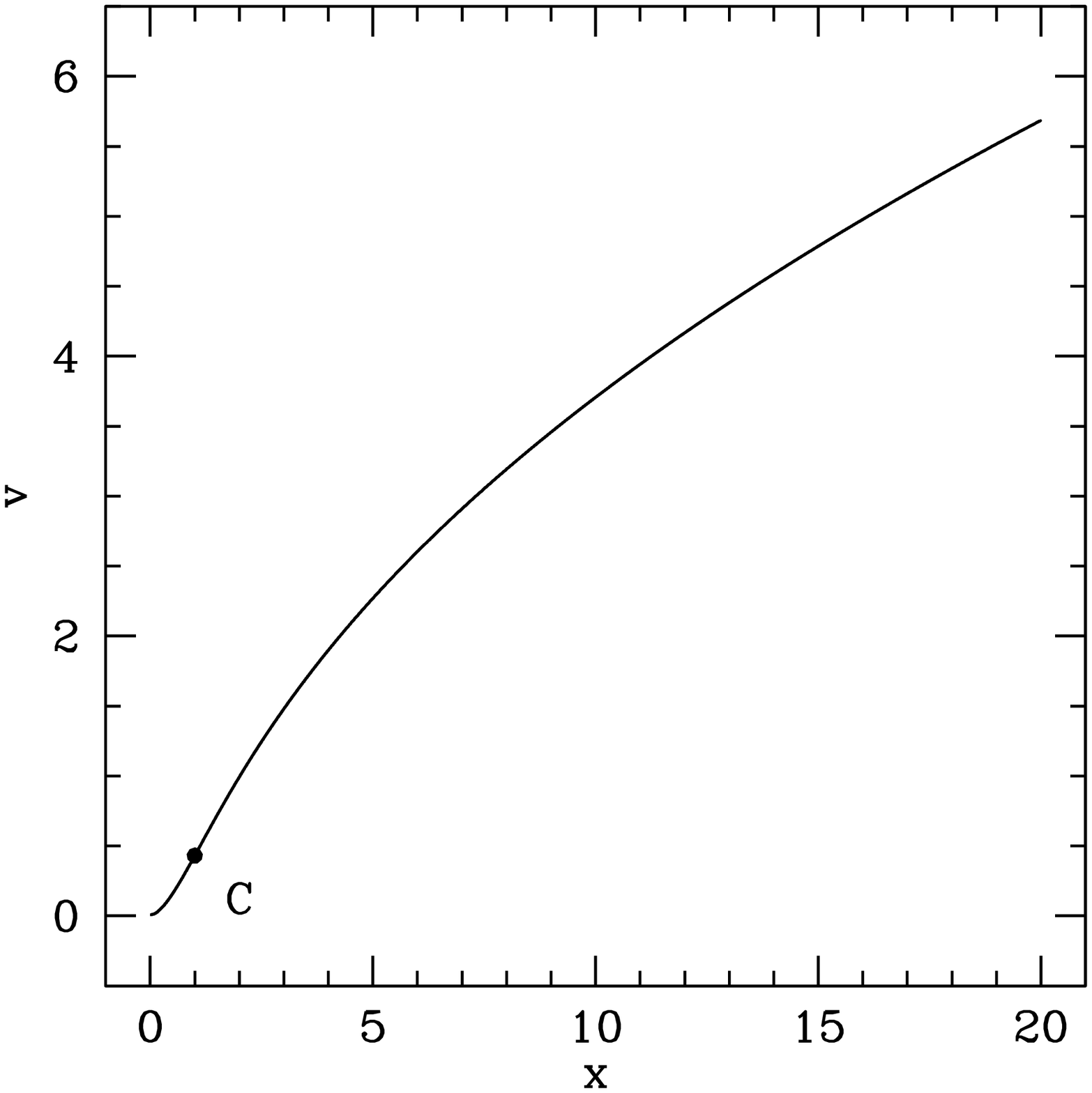}
\caption{
Velocity law for the FSH02 model ($s=10^{-4}$) obtained through
numerical integration of the equation of motion.
``C'' denotes the critical point.
}
\label{fig_fs02_va2}
\end{figure}

Figure~\ref{fig_fs02_va2} shows the velocity law of the FSH02 model for
$s=10^{-4}$.
One of the general results from the models we discuss in this paper is
that gas pressure effects produce only minor corrections in the
wind mass loss rate and in the velocity law;
this is illustrated here by comparing Figures~\ref{fig_fs02_vth}
and \ref{fig_fs02_va2}.

\begin{figure}
\epsscale{1.0}
\plotone{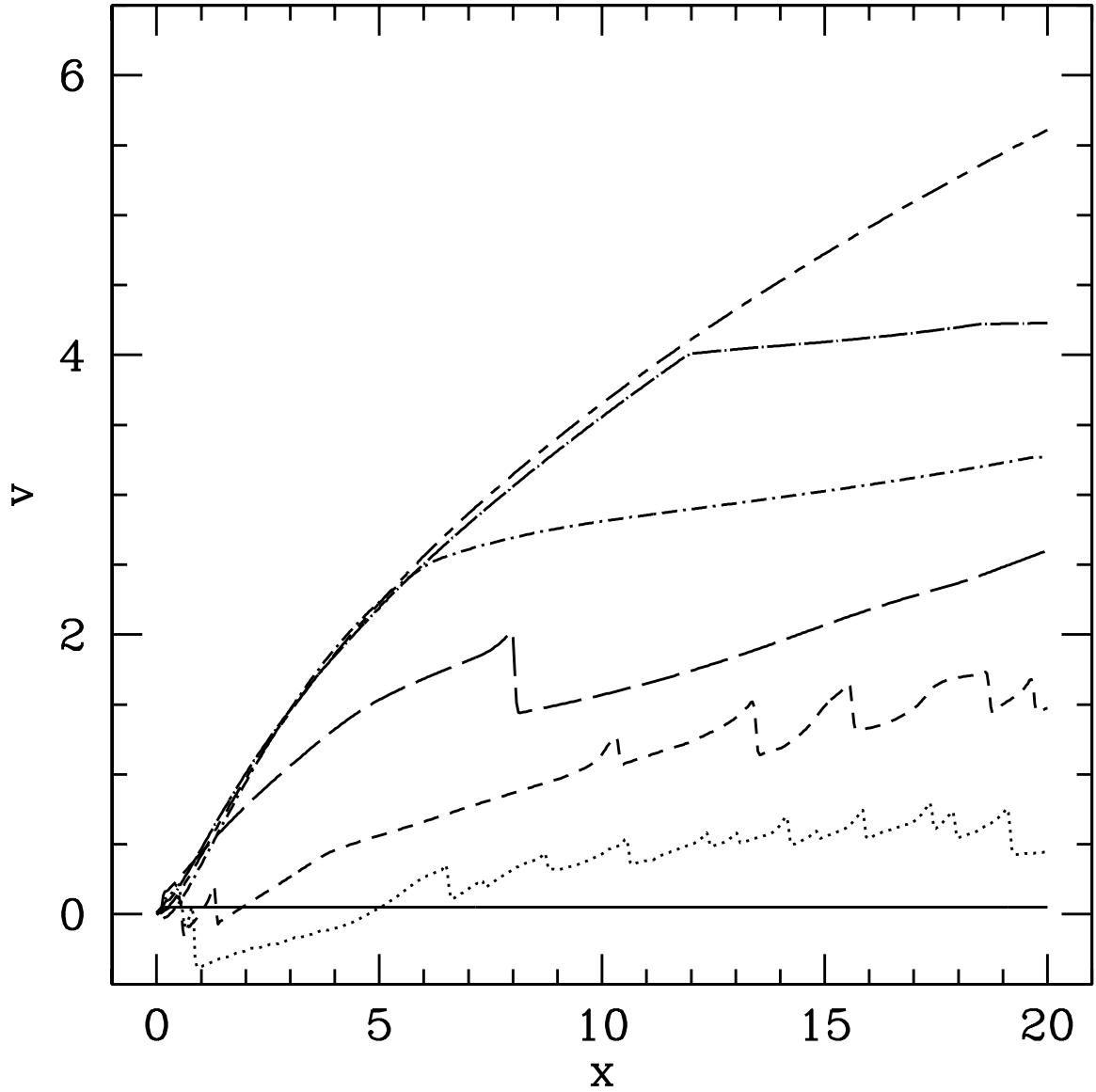}
\caption{
Time dependent velocity distribution for the FSH02 model
($s=10^{-4}$).
This plots shows that the steady solutions obtained through
the stationary numerical codes are stable.
In chronological order the plots are:
solid line,
dotted line,
short dashed line,
long dashed line,
short dot-dashed line,
long dot-dashed line,
and short dash-long dashed line.
}
\label{fig_fs02_num}
\end{figure}

Figure~\ref{fig_fs02_num} shows the results from the time-dependent
simulation. 
The solid line is the initial velocity distribution at
$t=0$;
the short dash-long dashed line is the velocity distribution found once
the code arrives at a steady state,
which is in excellent agreement with the steady state solution found
through the stationary codes.
This figure shows that the steady solutions found through the
stationary codes are stable.

Also,
we wish to note here that care must be taken when numerically
implementing the boundary conditions for the time-dependent
simulations.
As has been found for the case of line-driven stellar winds
\citep{owo94,cra96}
and for the case of CV line-driven disk winds
\citep{per00,per03},
by varying the numerical treatment of the boundary conditions,
one may be lead to obtain unsteady flows which are numerical artifacts
rather than intrinsic physical properties.
A similar situation is found for the time-dependent simulations of
MHD disk winds
(rather than line-driven disk winds)
\citep{ust99}
in which nonstationary flow may be artifacts of initial conditions
rather than being an intrinsic physical property of the flow.
The numerical time-dependent hydrodynamical models are based on the 
PPM numerical scheme
\citep{col84}.

A significant result from the analysis of the FSH02 model is that an
increase in gravity along streamlines does not imply an unsteady wind,
as is clearly shown in
Figure~\ref{fig_fs02_num}.

\section{The Standard ``Simple'' Models}
\label{simple_models}

As we discussed in the introduction,
one of the aims of this work is to study simple models that illustrate
the physics behind steady disk wind solutions without the elaborate
mathematical calculations required by more realistic models.
We proceed with this here.

The equation of motion for the standard models of this paper are 
given by equation~(\ref{equ_motion_w}).
Each model is defined by specifying expressions for the normalized body
force
[$g(x)$; gravitational plus continuum radiation acceleration],
the normalized area [$a(x)$],
and the normalized line opacity weighted flux [$f(x)$].

The three standard models we consider have the same expressions
for
$g(x)$
and for
$a(x)$,
and differ only in the expression for
$f(x)$.
The
$g(x)$
and
$a(x)$
for the standard models are given by

\begin{equation}
\label{equ_S_g}
  g(x)
=
  {x \over (1+x^2)^{3 \over 2}}
\;\;\;\; 
\end{equation}

\noindent
and 

\begin{equation}
\label{equ_S_a}
  a(x)
=
  1+x^2
\;\;\;\; .
\end{equation}

Equation~(\ref{equ_S_g}) is the exact expression for the vertical
component of the gravitational field of a compact object at disk
center,
where
$x=z/r_0$
and
$r_0$
is the radius of the streamline at wind base.
Equation~(\ref{equ_S_a}) corresponds to the geometry of the 1D disk
wind models of
\cite{per97b}.
The motive for selecting this area function,
in addition to it being a simple expression,
is that it has the correct asymptotic behavior as
$x \to 0$
(constant area: corresponding to parallel streamlines at the wind base)
and as
$x \to \infty$
($a \sim x^2$: corresponding to the divergence
               of streamlines under spherical symmetry).

For simplicity,
all standard models are assumed to have $\alpha = 1/2$~.

\subsection{The S Model}

\begin{figure}
\epsscale{1.0}
\plottwo{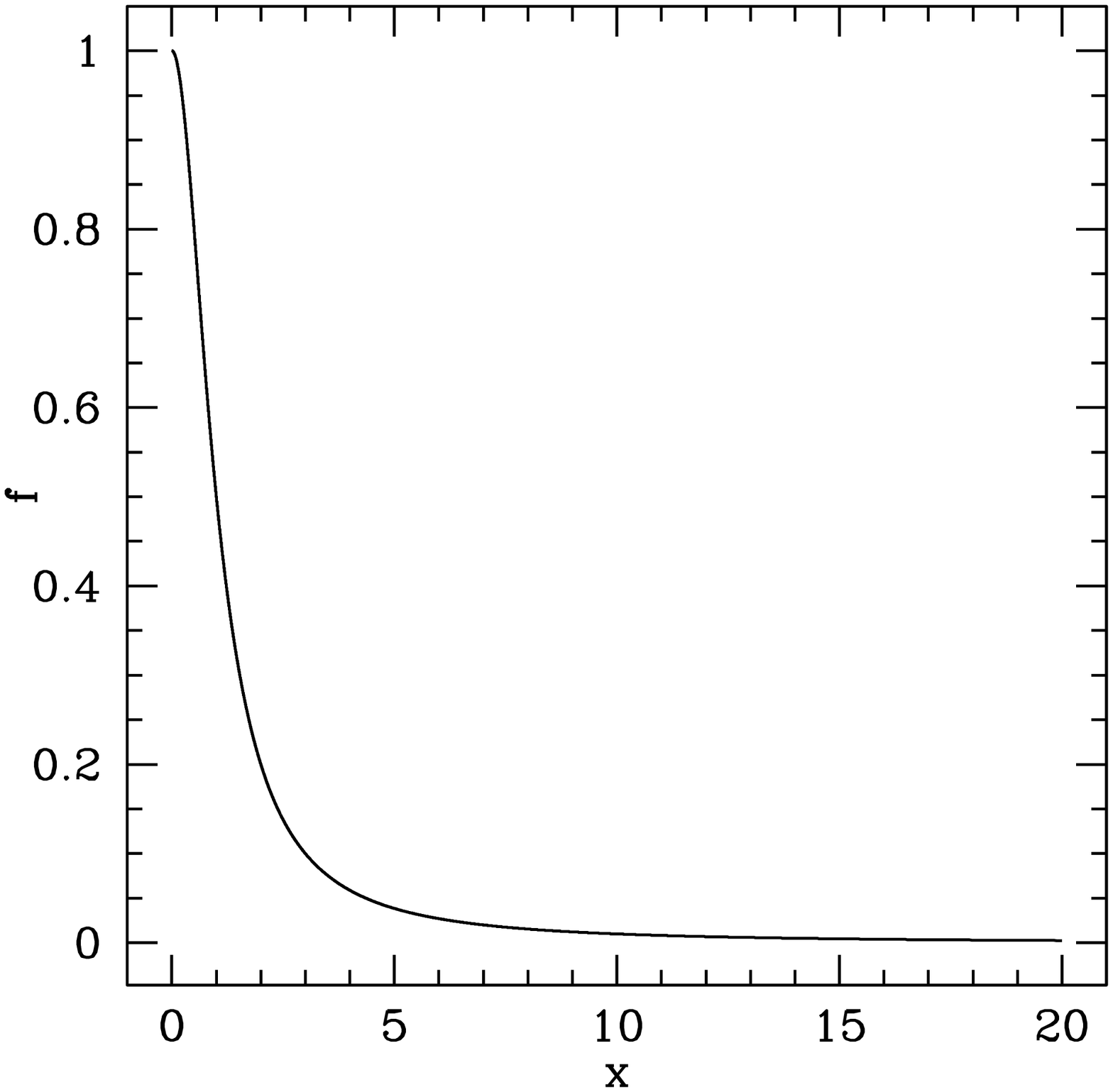}{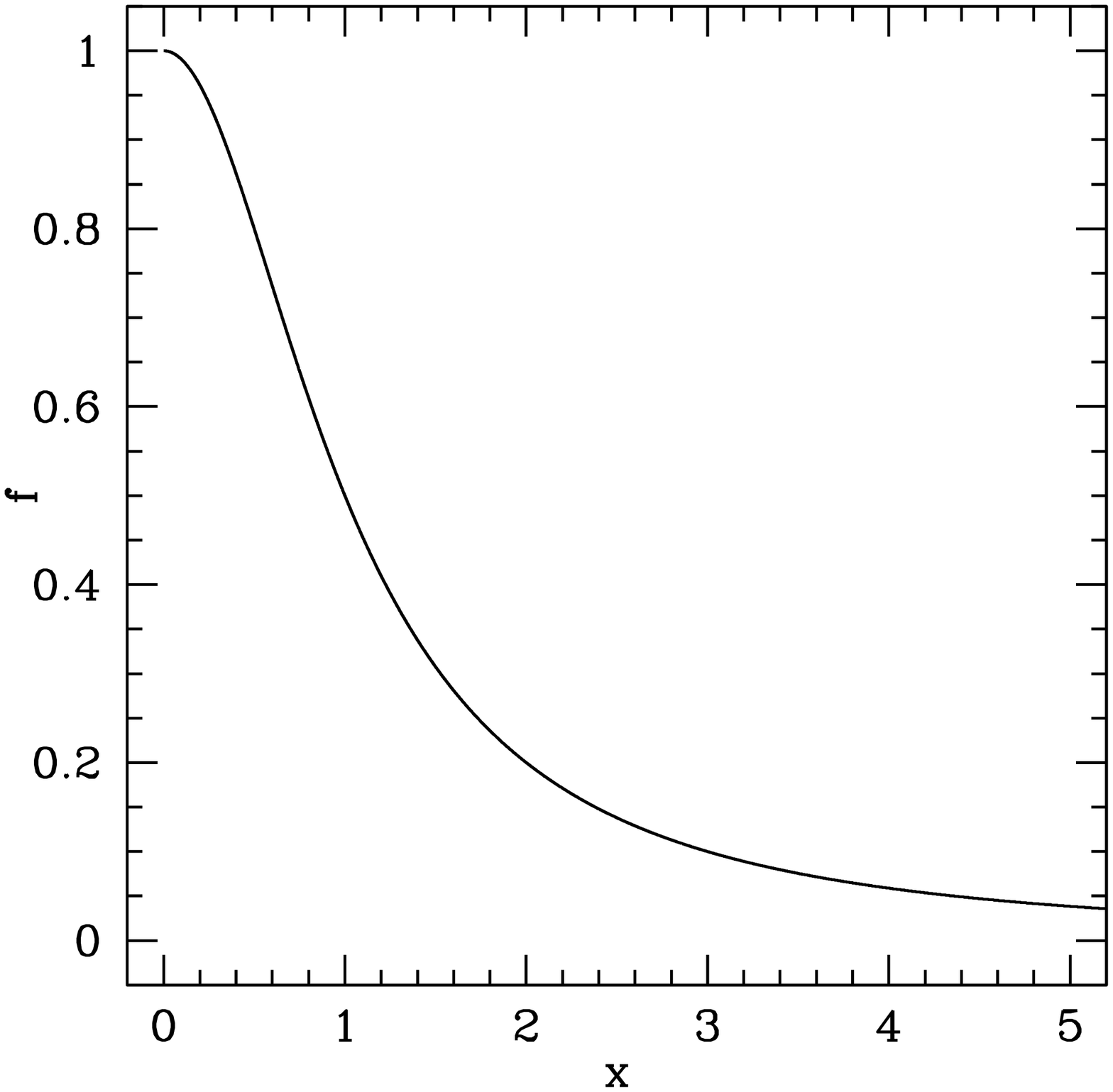}
\caption{
Line opacity weighted flux $f$ for the S~model.
The analytical form of the $f$ function for this model
(equation~[\ref{equ_S_f}]) was selected because
(in addition to its simplicity)
it has the correct asymptotic behavior as $x \to 0$
(constant flux near the disk surface)
and as $x \to \infty$
($f \sim x^{-2}$).
}
\label{fig_S_f}
\end{figure}

The standard ``S''~model is defined through the expression
(Figure~\ref{fig_S_f})

\begin{equation}
\label{equ_S_f}
  f(x)
=
  {1 \over 1+x^2}
\;\;\;\; .
\end{equation}

\noindent
The motive for selecting this expression for 
$f(x)$
is that it has the correct asymptotic behavior as
$x \to 0$
(constant flux near the surface)
and as
$x \to \infty$
($f \sim x^{-2}$).

\begin{figure}
\epsscale{1.0}
\plottwo{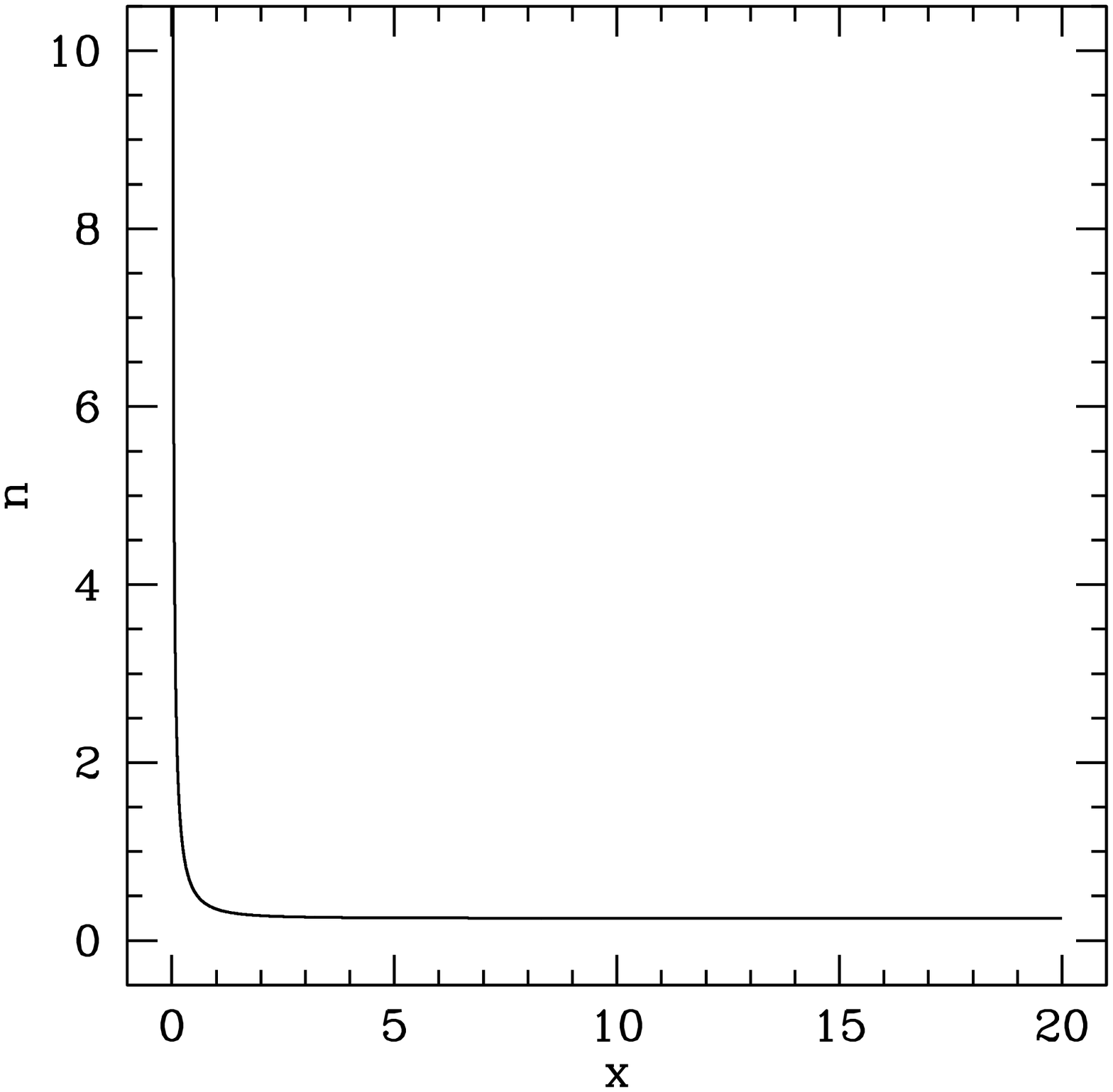}{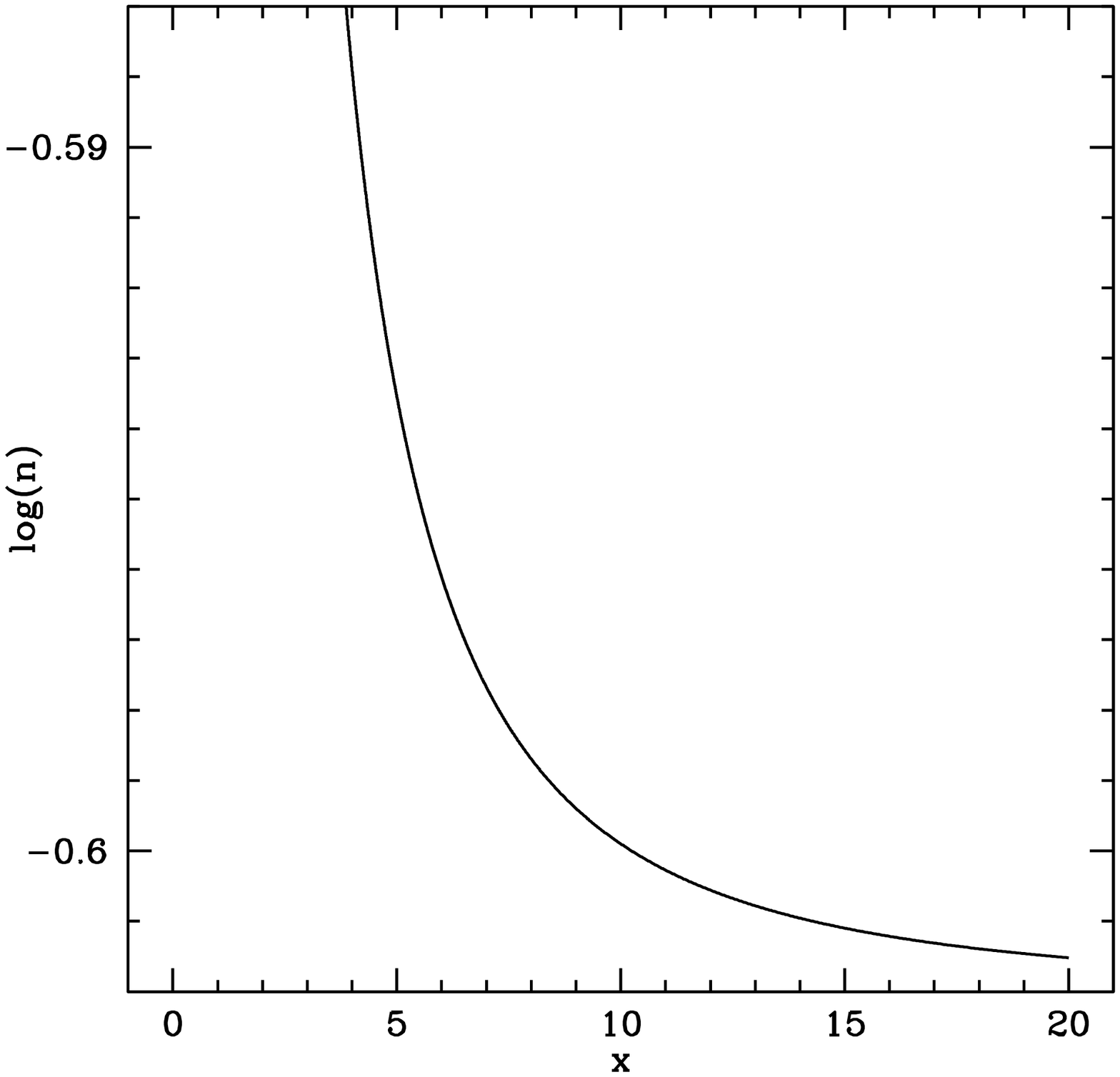}
\caption{
Nozzle function for the S model ($s=0$).
Note that the nozzle function is monotonically decreasing
throughout the spatial grid.
For this model (with $s=0$)
the nozzle function decreases monotonically to infinity,
with a finite values at ``infinity'' of $n_\infty = 1/4$.
}
\label{fig_S_n}
\end{figure}

In Figure~\ref{fig_S_n} we present the nozzle function $n(x)$ of the
S~model
(with $s=0$,
 neglecting gas pressure).
For this model the nozzle function is a monotonically decreasing
function with a finite value at infinity of
$n_\infty = 1/4$.
The critical point is thus at infinity.

\begin{figure}
\epsscale{1.0}
\plotone{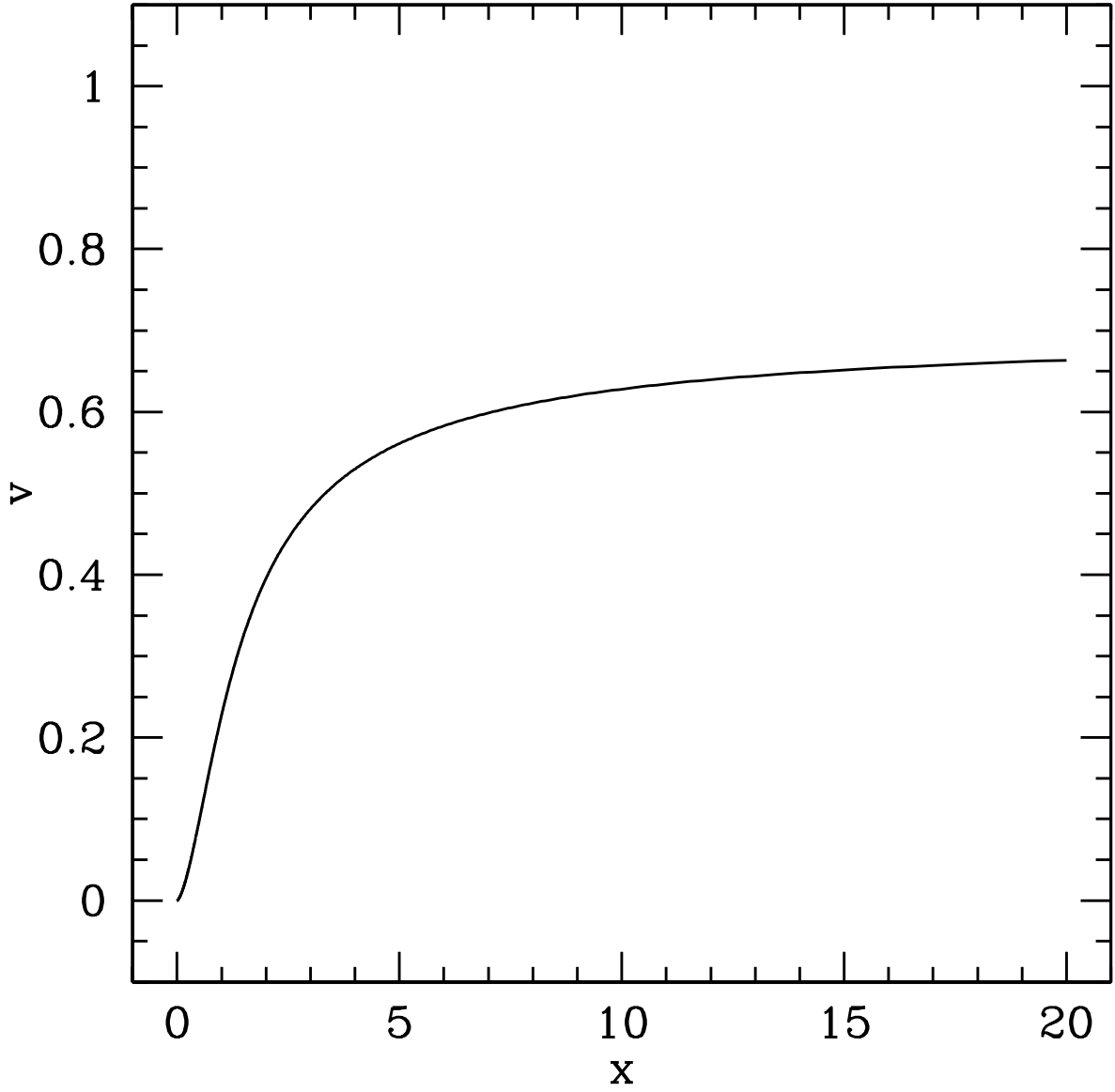}
\caption{
Velocity law for the S~model ($s=0$) obtained through
numerical integration of the equation of motion.
In the integration the critical point is assumed to be at
the highest grid spatial point ($x=20$).
}
\label{fig_S_v}
\end{figure}

In Figure~\ref{fig_S_v} we present the velocity law of the S~model
derived through numerical integration starting from the critical point.
For the S~model,
when $s=0$,
we assume that the critical point is at the highest spatial grid point
($x=20$ for the results presented here).

\begin{figure}
\epsscale{1.0}
\plotone{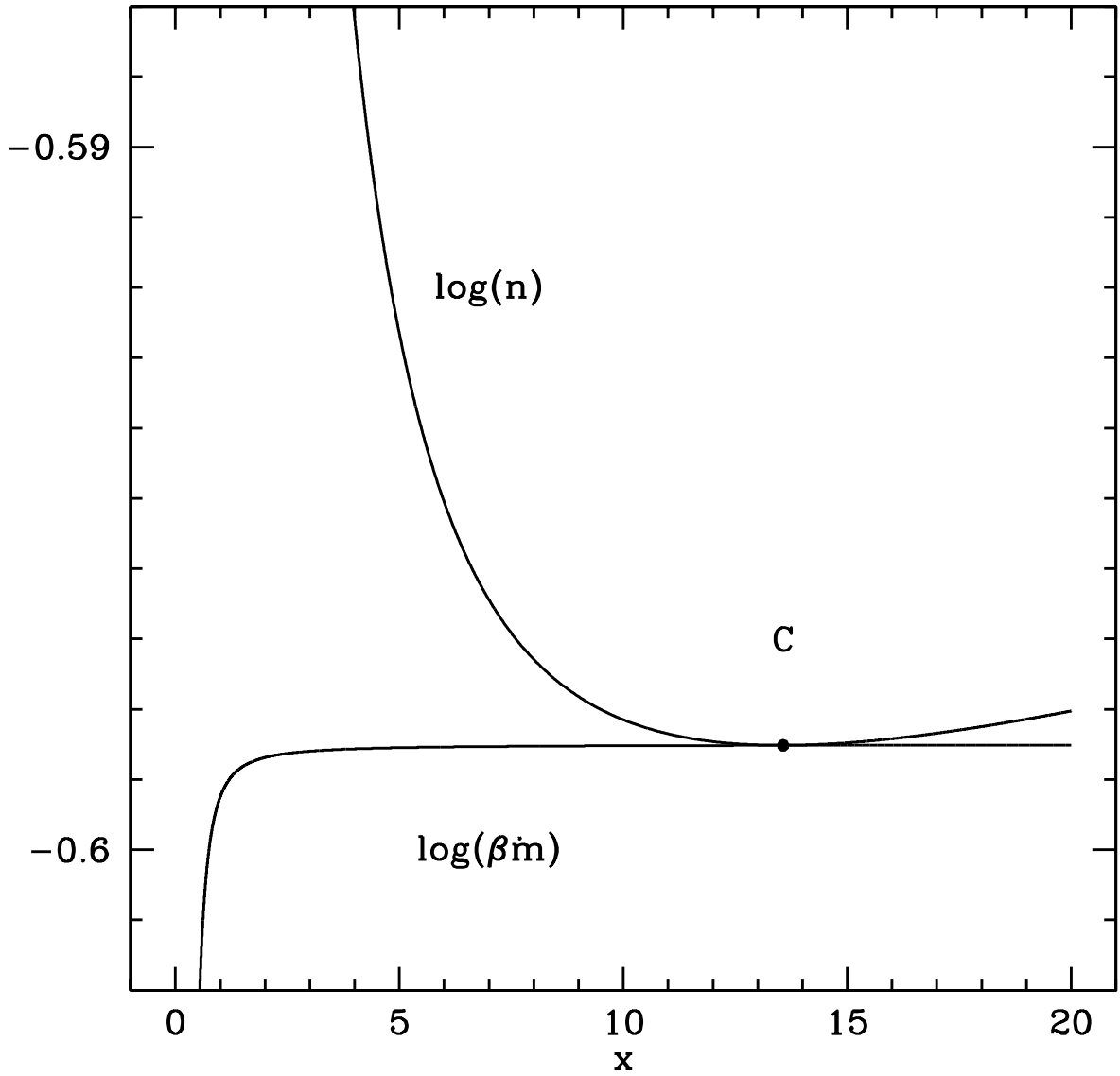}
\caption{
Logarithm of the nozzle function $\log(n[x])$ and the logarithm of the
product $\log(\beta[\omega[x]]\dot{m})$
for the S~model ($s=10^{-4}$).
Both plots intersect at,
and only at,
the critical point.
Note that (cf. Figure~\ref{fig_S_n}) when gas pressure effects are
included ($s > 0$) the nozzle function has a minimum at a finite
distance
(rather than having a minimum at infinity; cf. Figure~\ref{fig_S_n}).
The critical point is denoted by ``C''.
}
\label{fig_S_b}
\end{figure}

\begin{figure}
\plotone{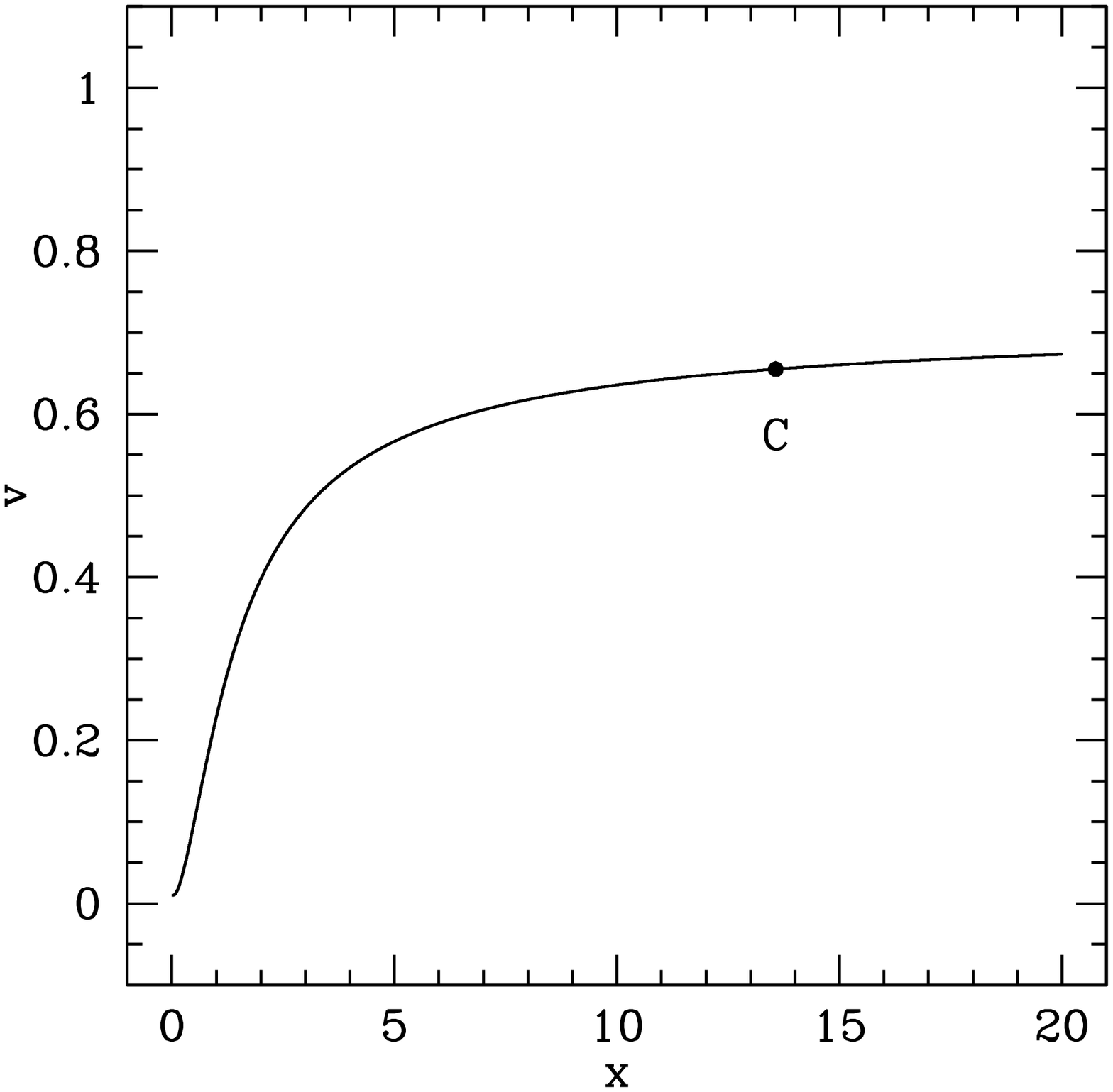}
\caption{
Velocity law for the S~model ($s=10^{-4}$) obtained through
numerical integration of the equation of motion.
``C'' denotes the critical point.
}
\label{fig_S_va2}
\end{figure}

\begin{figure}
\plotone{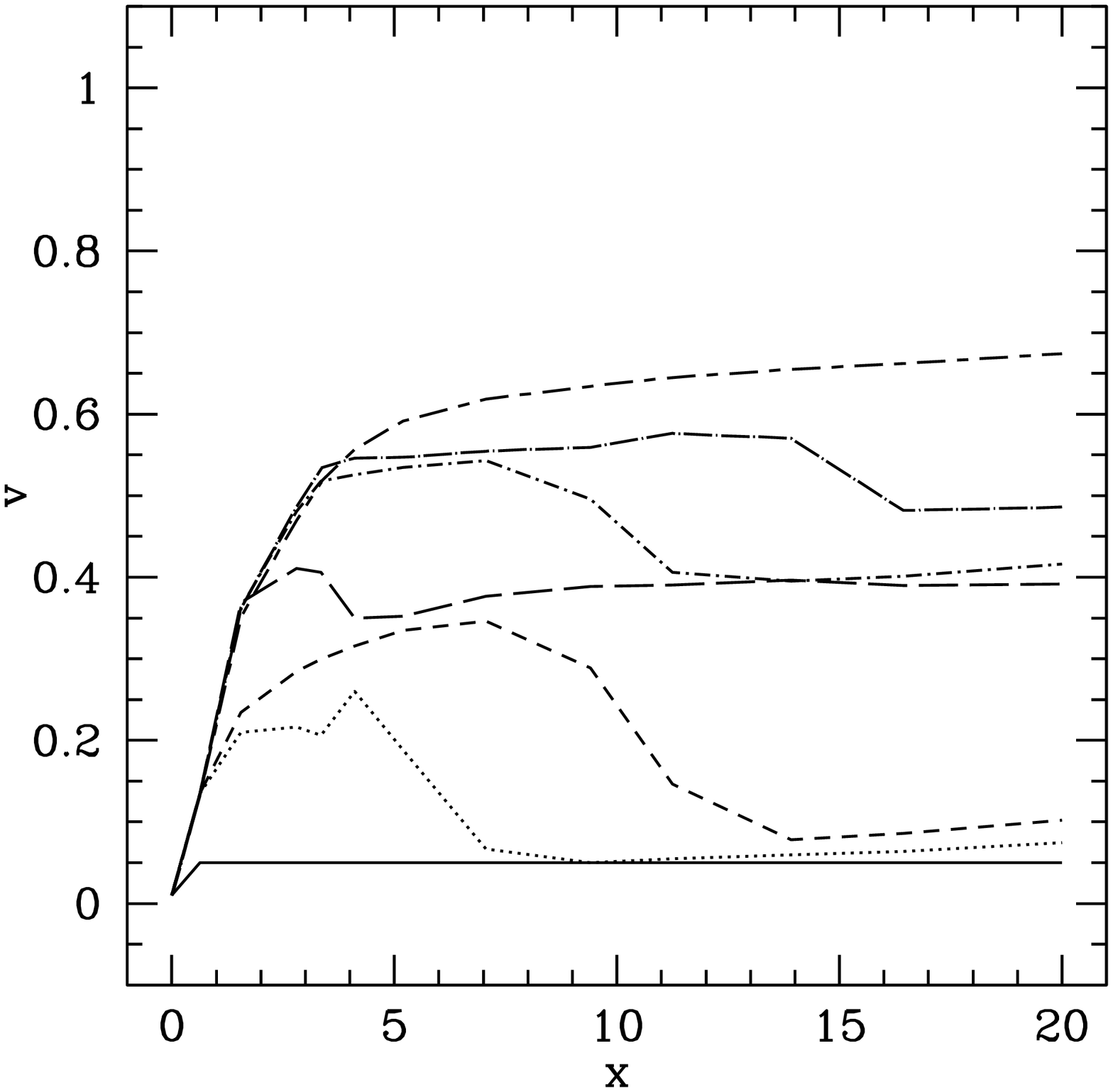}
\caption{
Time dependent velocity distribution for the S~model
($s=10^{-4}$).
This plots shows that the steady solutions obtained through
the stationary numerical codes are stable.
In chronological order the plots are:
solid line,
dotted line,
short dashed line,
long dashed line,
short dot-dashed line,
long dot-dashed line,
and short dash-long dashed line.
The code arrives at a steady state after a small multiple of the
``crossing time''
(the time it would take a particle in the wind to travel from the
 the disk surface [$x=0$] to the right end of the spatial grid
 [$x=20$] once the steady state is achieved).
}
\label{fig_S_num}
\end{figure}

Now,
as we have seen from a physical standpoint,
the wind mass loss rate is determined by the minimum of the nozzle
function.
When neglecting gas pressure,
the S~model places the critical point at infinity.
Thus,
although a well defined stationary solution can mathematically be
found,
we are still left with the physical difficulty of having to account for
the travel of information through an infinite distance
(i.e., from the wind base to the critical point)
in a finite time.
However,
once gas pressure effects are taken into account
($s>0$),
the nozzle function has a well defined minimum at a finite distance
from the wind base.
In turn,
the critical point is slightly to the right of the minimum,
also at a finite distance from the wind base.
Therefore,
although the results indicate that gas pressure effects do not
significantly affect the value of the wind mass loss rate or the wind
velocity law,
for the S~model gas pressure is important in accounting for the
existence of steady wind solutions,
which we have confirmed through numerical time-dependent hydrodynamic
codes.
To show this in Figure~\ref{fig_S_b},
the nozzle function
$n(x)$
is plotted simultaneously with the product
$\beta([\omega(x)]) \dot{m}$
of the S~model for
$s=10^{-4}$.
Figure~\ref{fig_S_va2} shows the velocity law of the S~model for
$s=10^{-4}$.
Our observation that gas pressure effects produce only minor
corrections in the wind mass loss rate and the velocity law is
illustrated by comparing Figures~\ref{fig_S_v} and \ref{fig_S_va2}.
Figure~\ref{fig_S_num} shows the results from the time-dependent
simulation for the S~model.
The solid line is the initial velocity distribution at
$t=0$.
The short dash-long dashed line is the velocity distribution found once
the code arrives at a steady state,
and this is in excellent agreement with the steady state solution found
through the stationary codes.
Since the time steady solution agrees with the stationary code,
the stationary code solutions are stable.

\subsection{The I Model}

The ``I''~model is a modified version of the S~model.
The difference with respect to the S~model is a subtle modification of
the
$f(x)$
function with the goal of mimicking the flux distribution of a standard
Shakura-Sunyaev disk
\citep{sha73}
in the inner disk region,
where the scale height of the flux originating from the disk is
slightly larger than the scale height of gravity for a compact mass at
the center of the disk
\citep[e.g.,][]{per97a,per00}.

\begin{figure}
\epsscale{1.0}
\plottwo{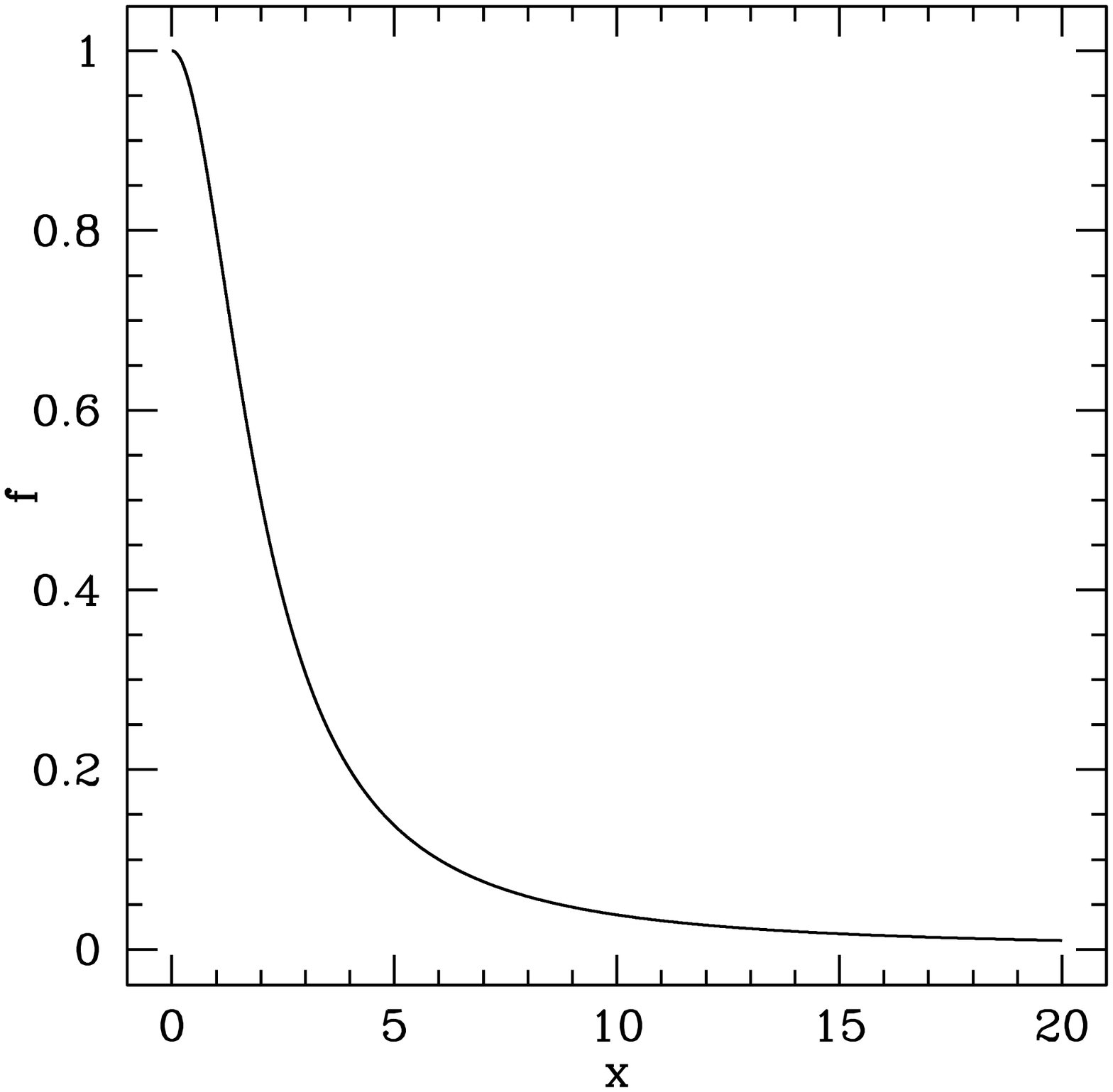}{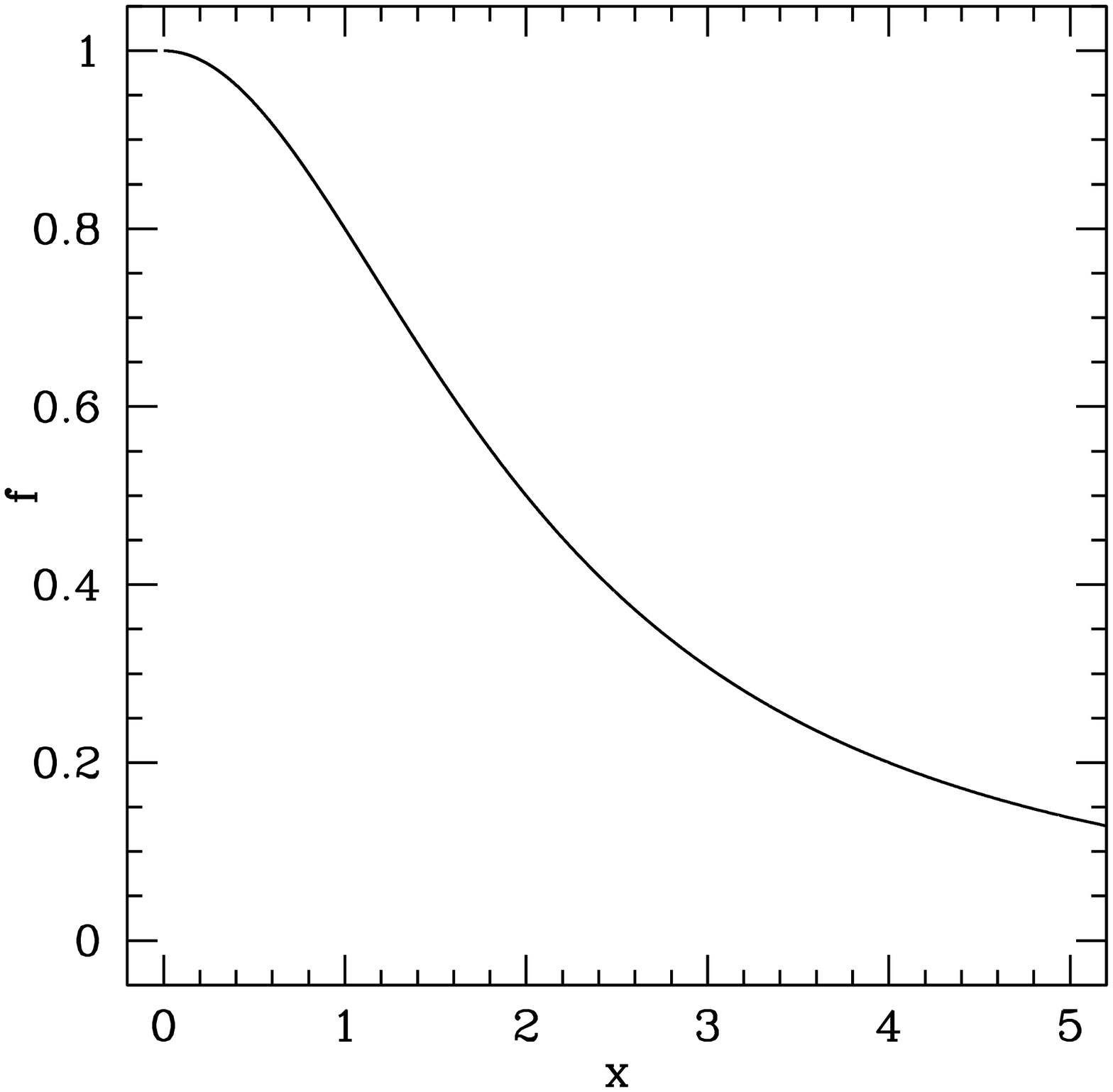}
\caption{
Line opacity weighted flux $f$ for the I~model.
The analytical form of the $f$ function for this model
(equation~[\ref{equ_I_f}])
was obtained by adding a subtle change to the $f$ function of the
S~model (equation~[\ref{equ_S_f}])
so as to produce an increase in the ``scale height'' of the flux.
In turn,
the motive for this is to mimic the inner disk region of a standard
Shakura-Sunyaev disk where the scale height of the flux is slightly
higher than the scale height of gravity for a compact object at disk
center
(cf. Figure~\ref{fig_S_f}).
}
\label{fig_I_f}
\end{figure}

The I~model is defined through the expression
(see Figure~\ref{fig_I_f})

\begin{equation}
\label{equ_I_f}
  f(x)
=
  {1 \over 1+\left(x \over 2\right)^2}
\;\;\;\; .
\end{equation}

\begin{figure}
\epsscale{1.0}
\plottwo{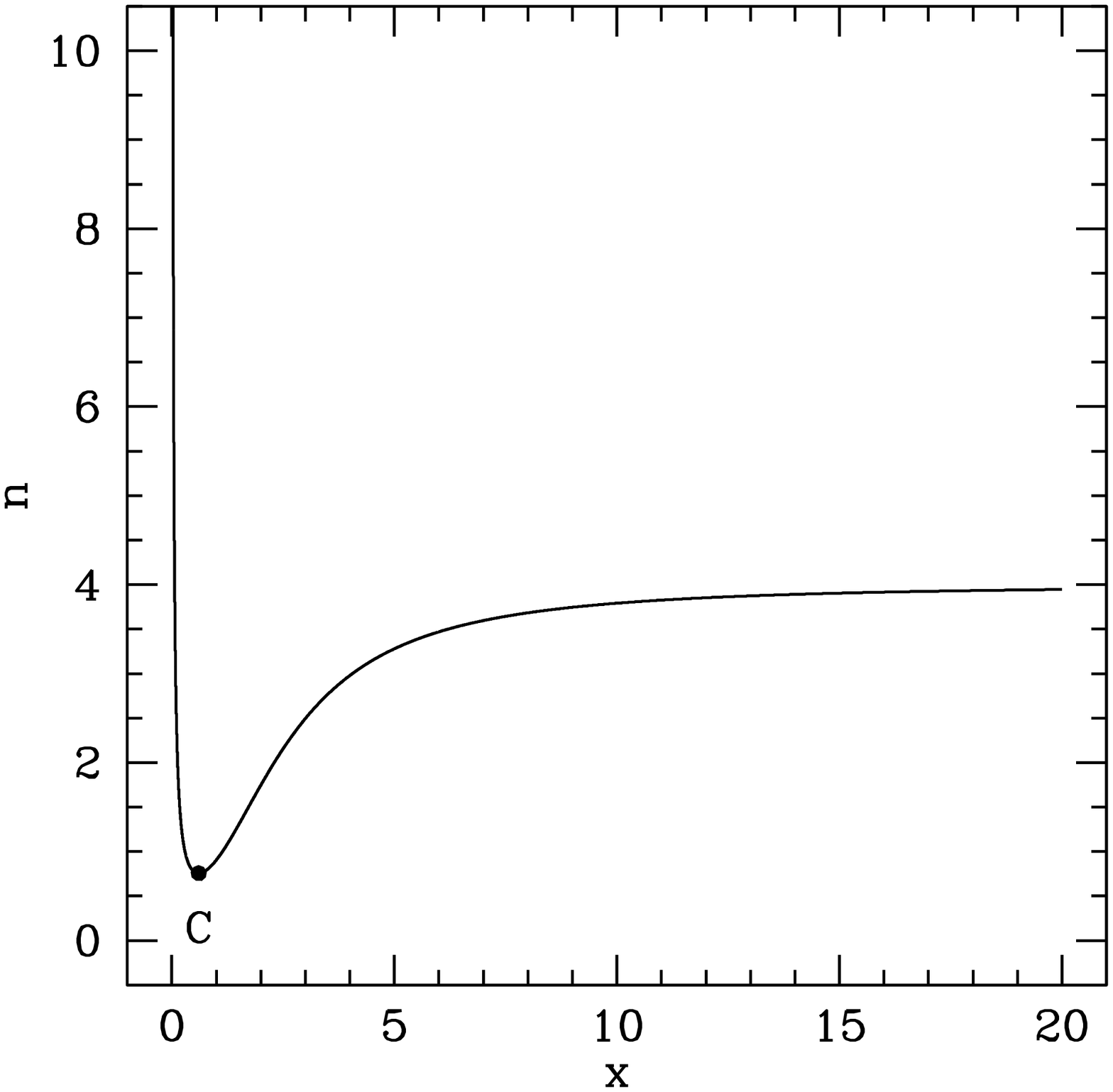}{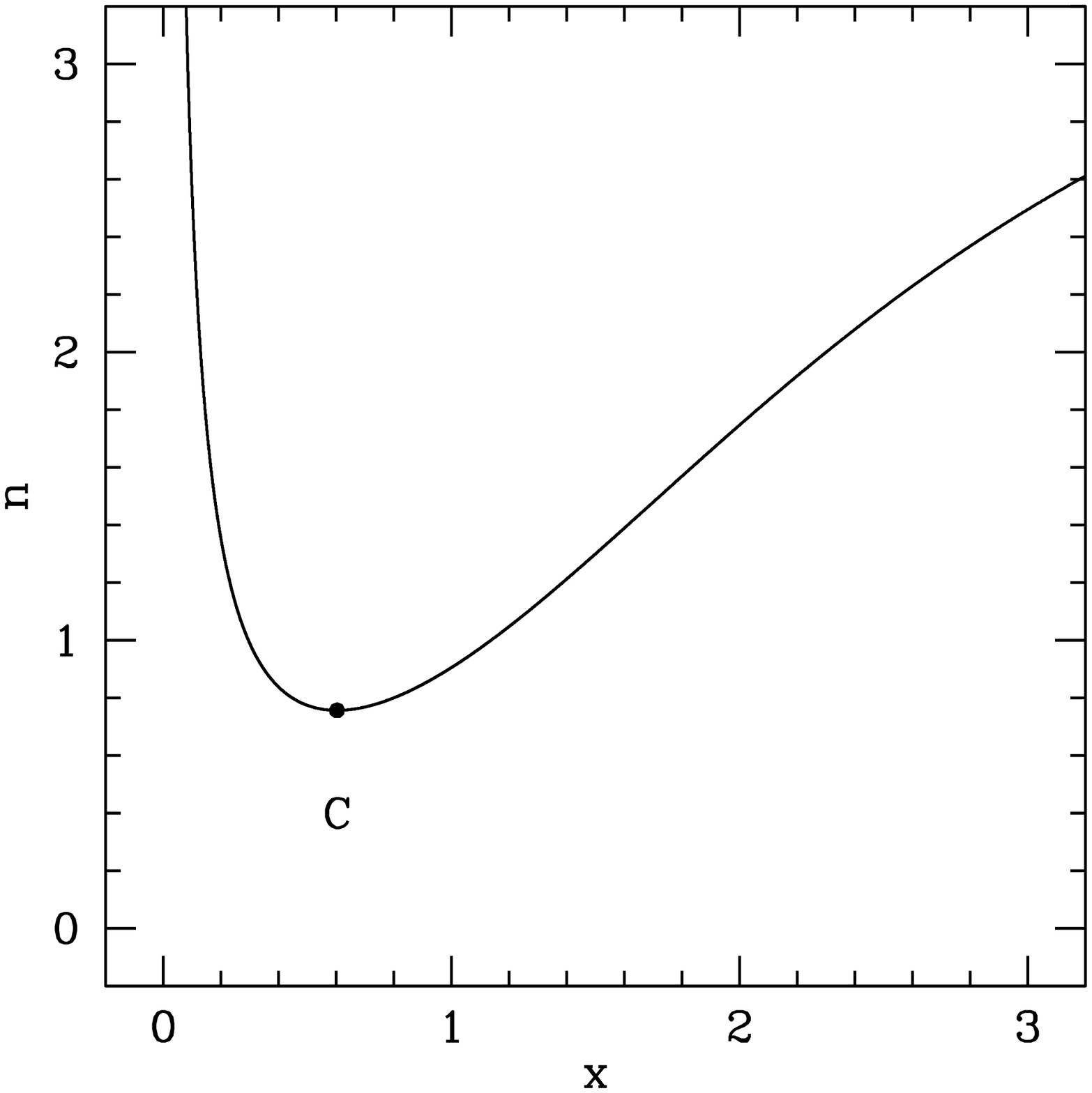}
\caption{
Nozzle function for the I~model ($s=0$).
When $s=0$ the critical point is determined by the position of the
minimum of the nozzle function.
``C'' denotes the critical point.
}
\label{fig_I_n}
\end{figure}

\begin{figure}
\epsscale{1.0}
\plotone{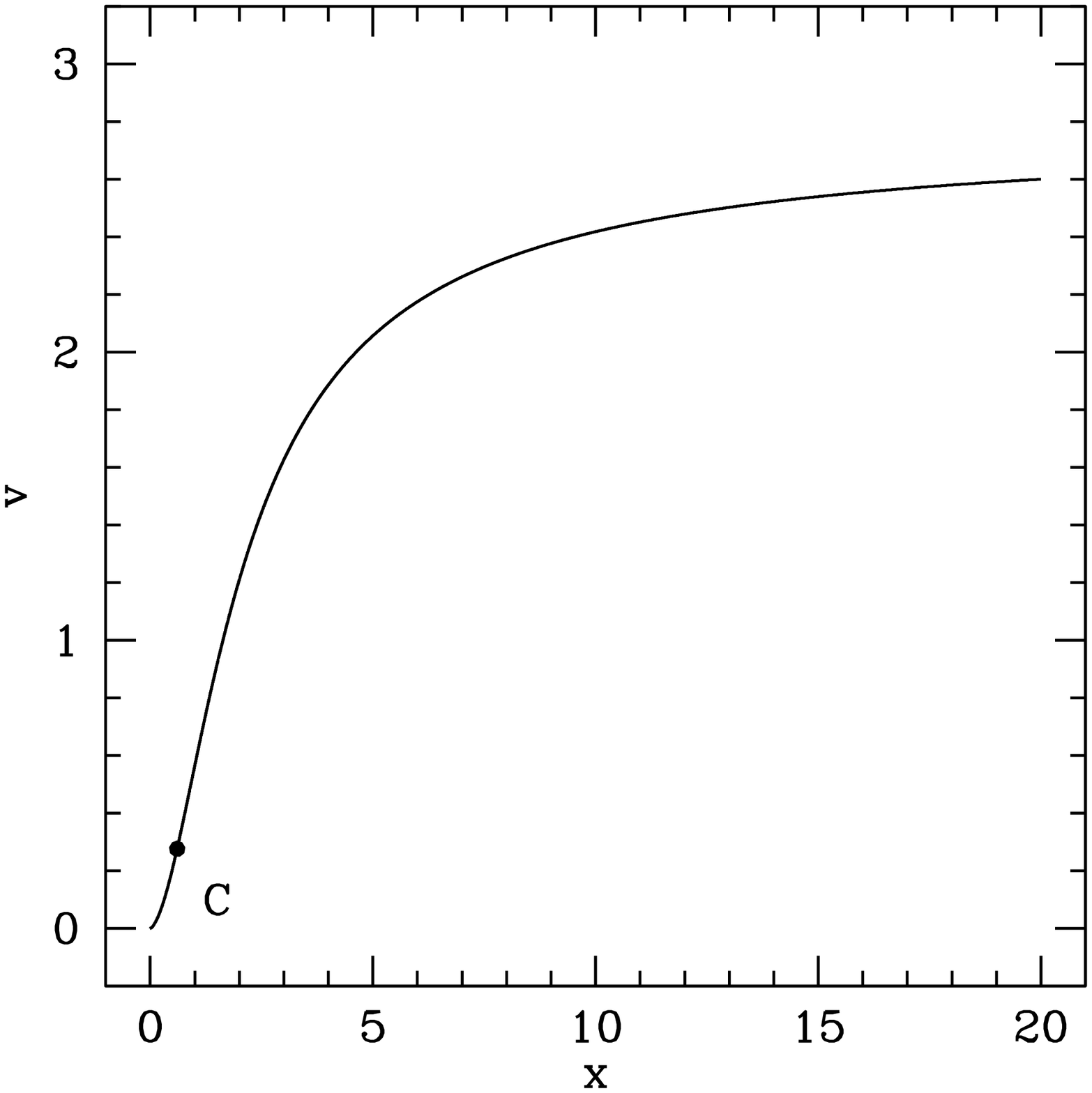}
\caption{
Velocity law for the I~model ($s=0$) obtained through numerical
integration of the equation of motion.
``C'' denotes the critical point.
}
\label{fig_I_v}
\end{figure}

\begin{figure}
\plottwo{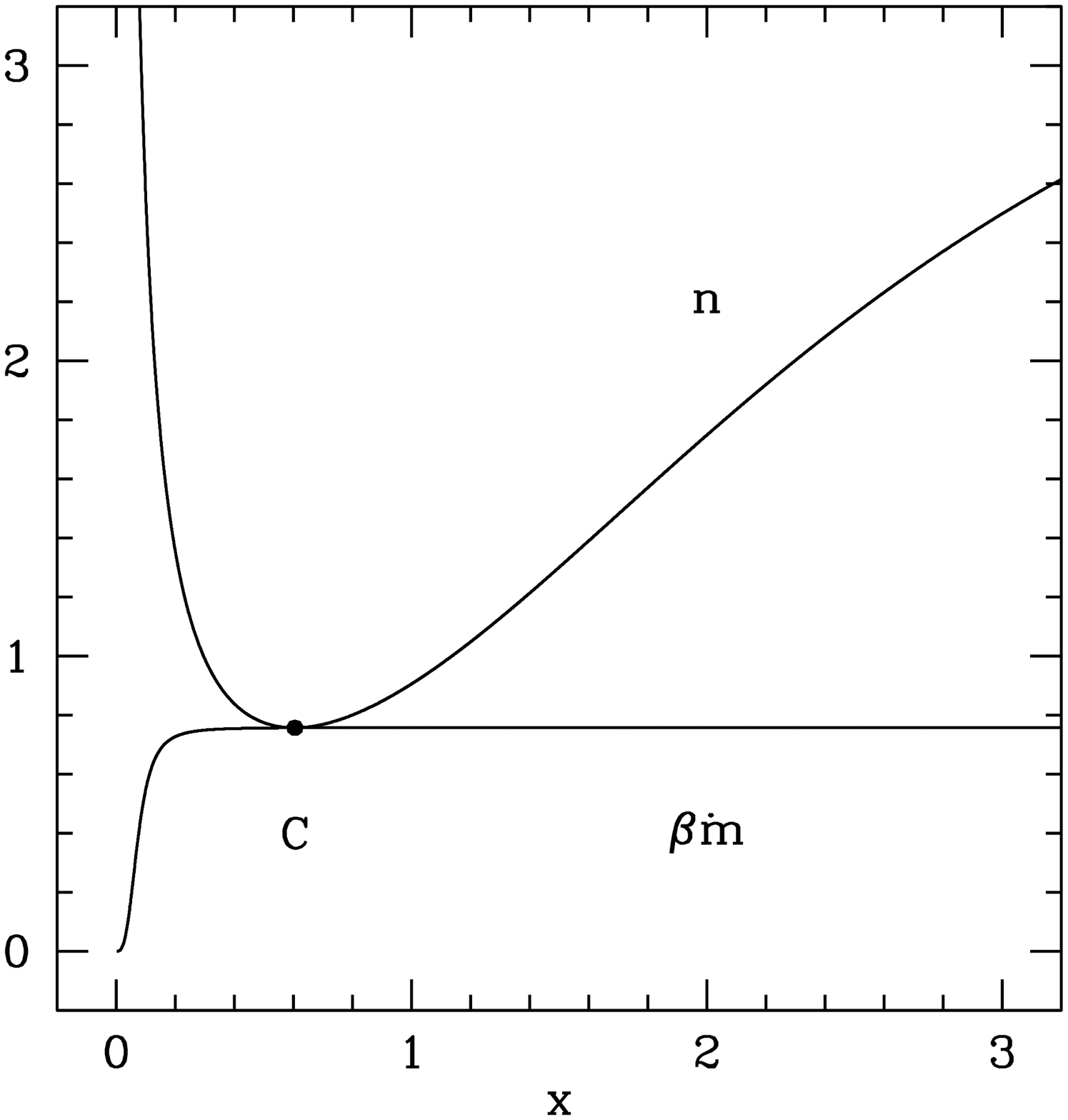}{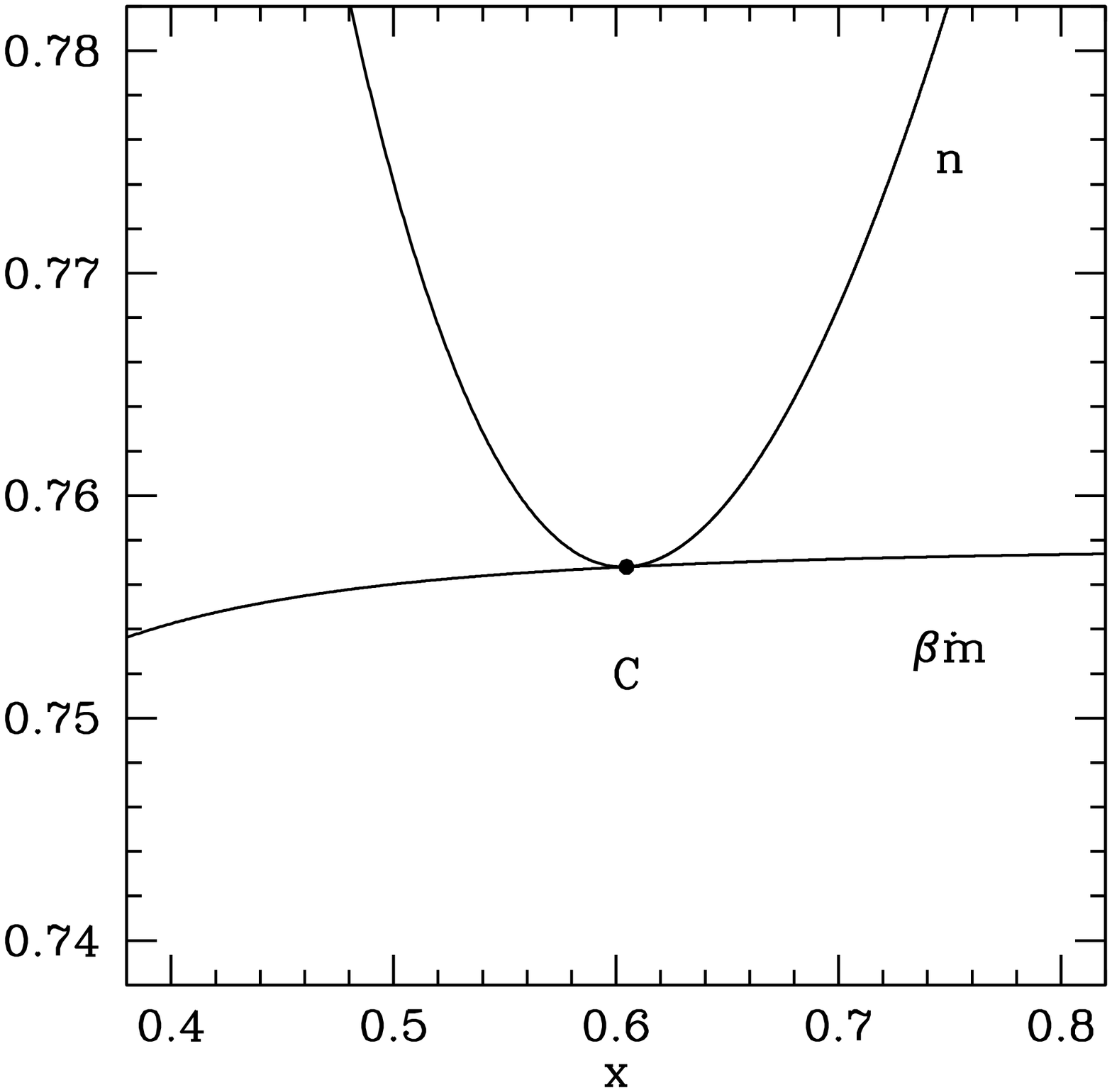}
\caption{
Nozzle function $n(x)$ and the product
$\beta(\omega[x]) \dot{m}$
for the I~model ($s=10^{-4}$).
Note that when gas pressure effects are included ($s>0$),
``$\beta(\omega[x])$'' is a monotonically increasing function
(equation~[\ref{equ_b}]);
thus (since $\dot{m}$ is a constant) it follows that the critical
point is shifted slight to the right of the nozzle minimum
(rather than {\it exactly} at the nozzle minimum,
 as when $s=0$).
``C'' denotes the critical point.
}
\label{fig_I_b}
\end{figure}

\begin{figure}
\plotone{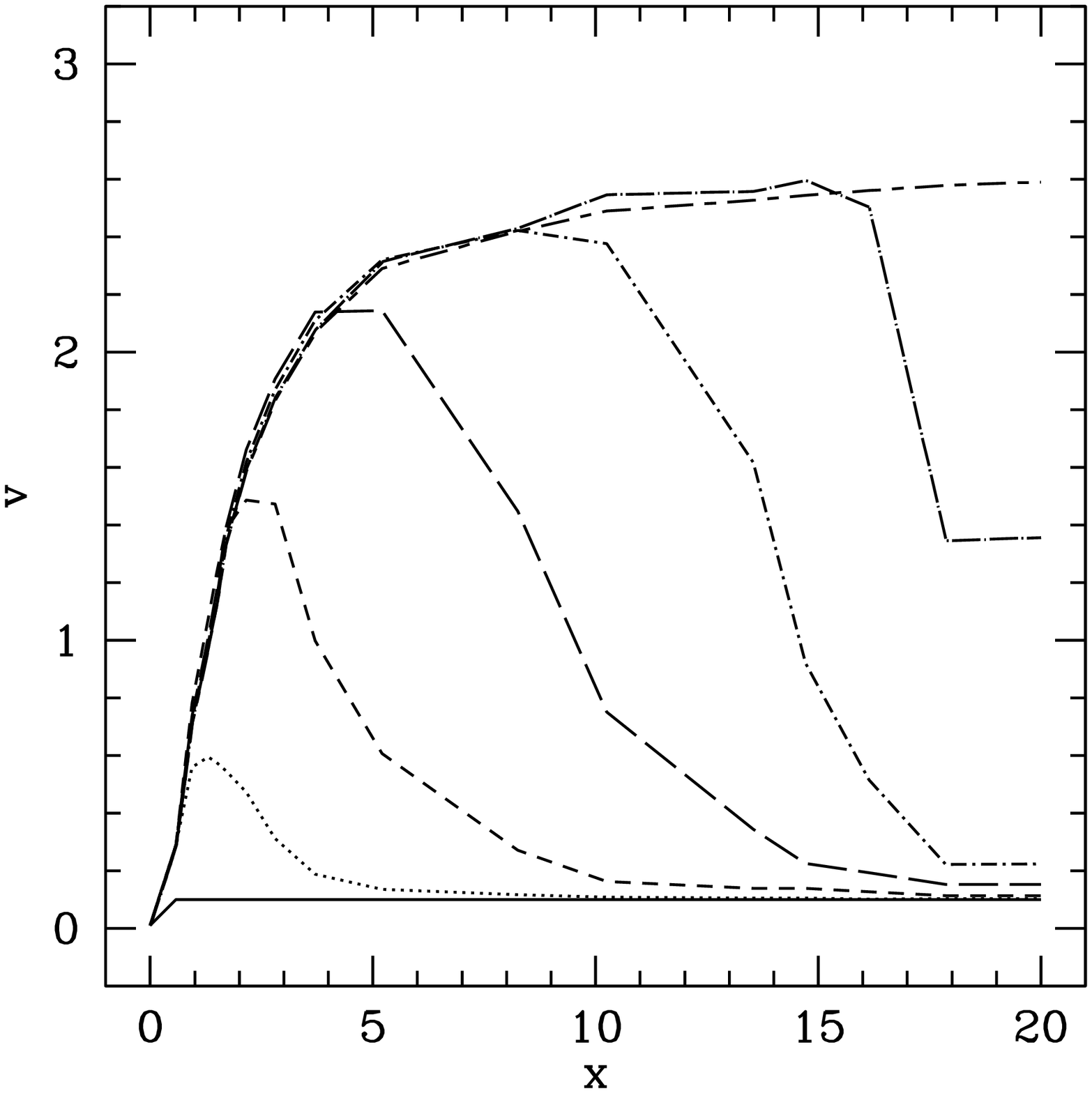}
\caption{
Time dependent velocity distribution for the I~model
($s=10^{-4}$).
This plots shows that the steady solutions obtained through
the stationary numerical codes are stable.
In chronological order the plots are:
solid line,
dotted line,
short dashed line,
long dashed line,
short dot-dashed line,
long dot-dashed line,
and short dash-long dashed line.
The code arrives at a steady state after a small multiple of the
``crossing time''
(the time it would take a particle in the wind to travel from the
 the disk surface [$x=0$] to the right end of the spatial grid
 [$x=20$] once the steady state is achieved).
}
\label{fig_I_num}
\end{figure}

In Figure~\ref{fig_I_n} we present the nozzle function
$n(x)$
of the I~model (with $s=0$).
The critical point is determined by the minimum of $n(x)$.
In Figure~\ref{fig_I_v} we present the velocity law of the I~model
derived through numerical integration starting from the critical point.
We have also computed the nozzle function and the velocity law of the
I~model for
$s=10^{-4}$,
but the corresponding corrections due to gas pressure are not
significant,
so we do not show the plots here to avoid redundancy.
Figure~\ref{fig_I_b} shows the nozzle function $n(x)$ plotted
simultaneously with the product
$\beta([\omega(x)]) \dot{m}$
for
$s=10^{-4}$.
These two graphs intersect at,
and only at,
the critical point.
Figure~\ref{fig_I_num} shows the results from the time-dependent
simulation for the I~model.
The solid line is the initial velocity distribution at $t=0$;
the short dash-long dashed line is the velocity distribution found once
the code arrives at a steady state.
Again we see that the time-dependent code converges and the results are
in excellent agreement with the steady state solution found through
stationary codes.
Thus,
this shows that the steady solutions found through stationary codes are
stable.

\subsection{The O Model}

As with the I~model, the ``O''~model is a modified version of the
S~model.
The difference between the O~model and the S~model is a subtle
modification of the
$f(x)$
function,
but now with the goal of mimicking the flux distribution of a standard
disk in the outer region of the disk.
In the outer region of a standard disk
\citep{sha73}
the flux initially increases with height from the disk surface
\citep[e.g.,][]{per97a,per00}.
This increase is due to flux emanating from interior radii.

\begin{figure}
\epsscale{1.0}
\plottwo{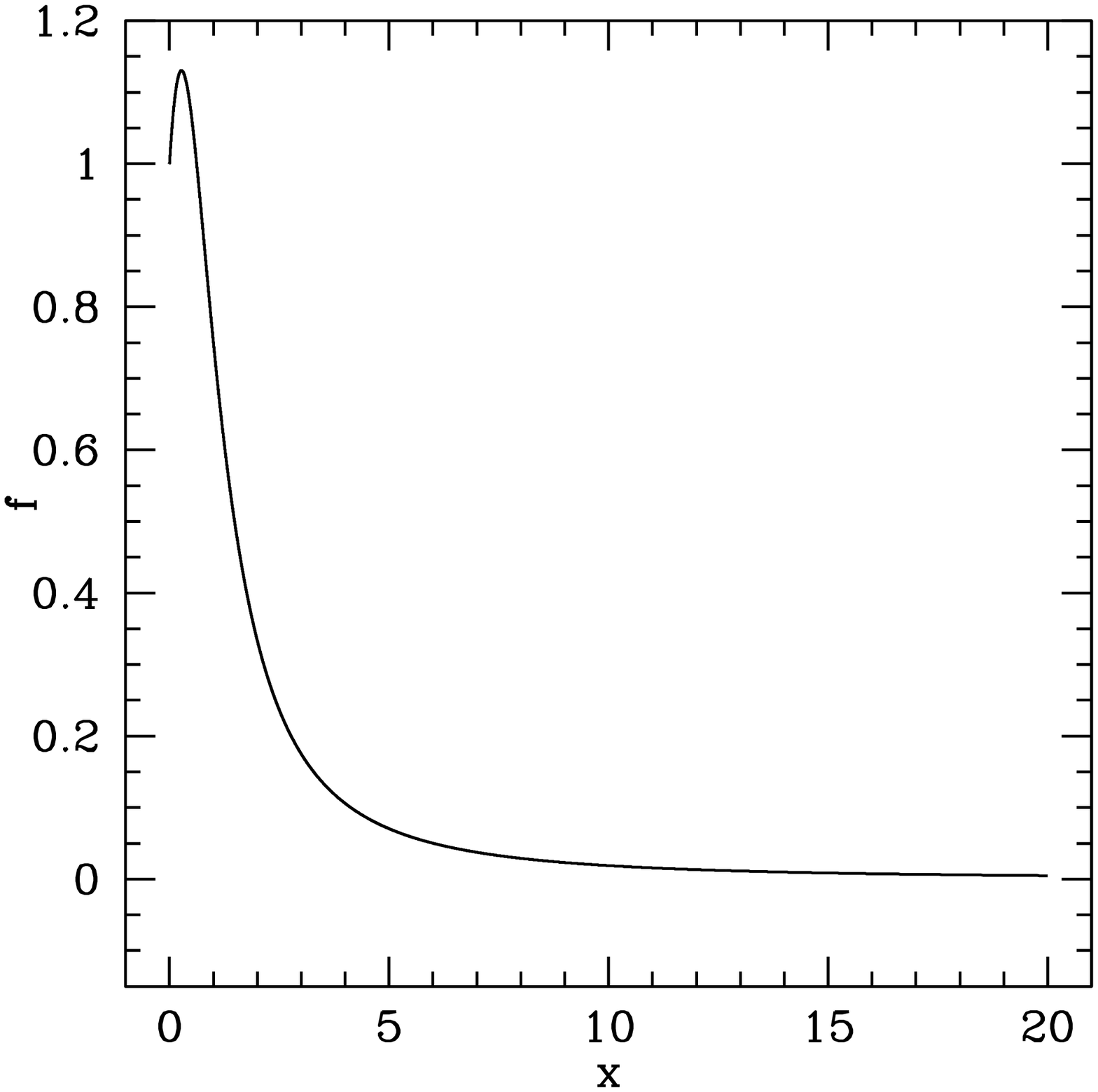}{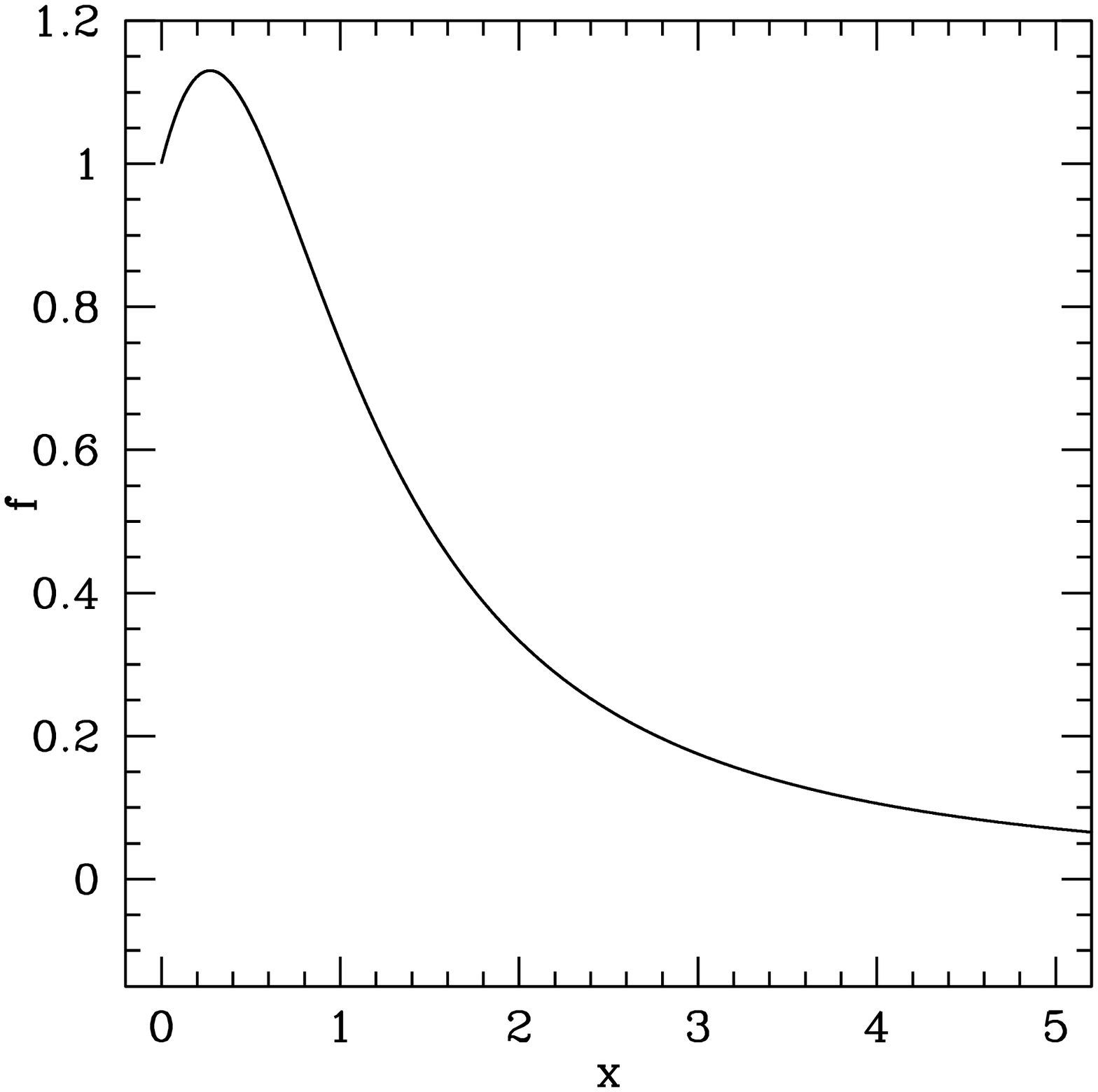}
\caption{
Line opacity weighted flux $f$ for the O~model.
The analytical form of the $f$ function for this model
(equation~[\ref{equ_O_f}])
was obtained by adding a subtle change to the $f$ function of the
S~model (equation~[\ref{equ_S_f}])
so as to produce a flux distribution with an initially increasing
flux from the disk surface and a subsequent local maximum.
In turn,
the motive for this is to mimic the outer disk region
(rather than the inner disk region as in Figure~\ref{fig_I_f}),
of a standard Shakura-Sunyaev disk,
where the vertical radiation flux initially increases with height
from the disk surface
(cf. Figure~\ref{fig_S_f}).
}
\label{fig_O_f}
\end{figure}

The O~model is defined through the expression
(see Figure~\ref{fig_O_f})

\begin{equation}
\label{equ_O_f}
  f(x)
=
  {1 + {x \over 1 + x} \over 1+x^2}
\;\;\;\; .
\end{equation}

\noindent

\begin{figure}
\epsscale{1.0}
\plottwo{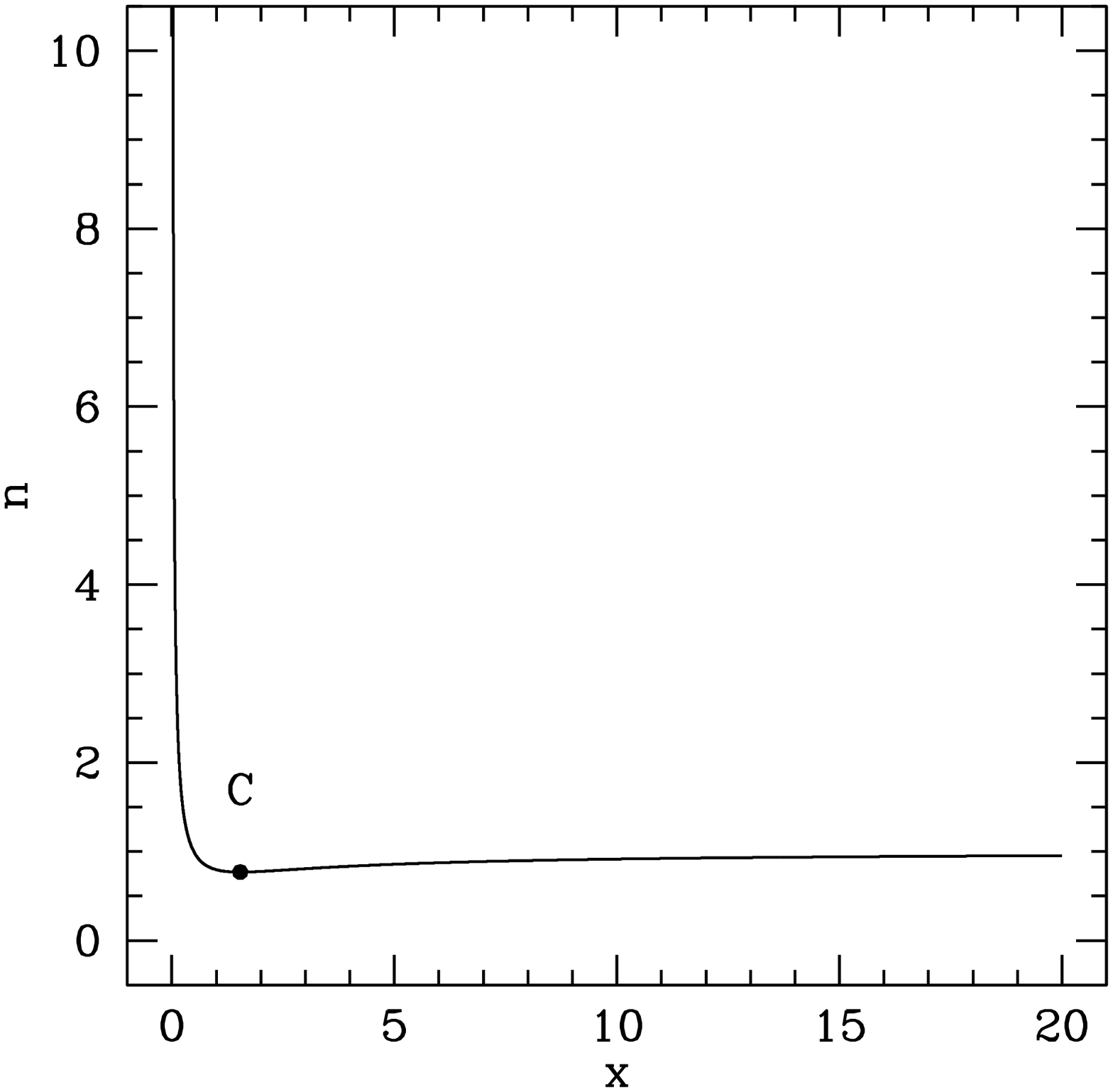}{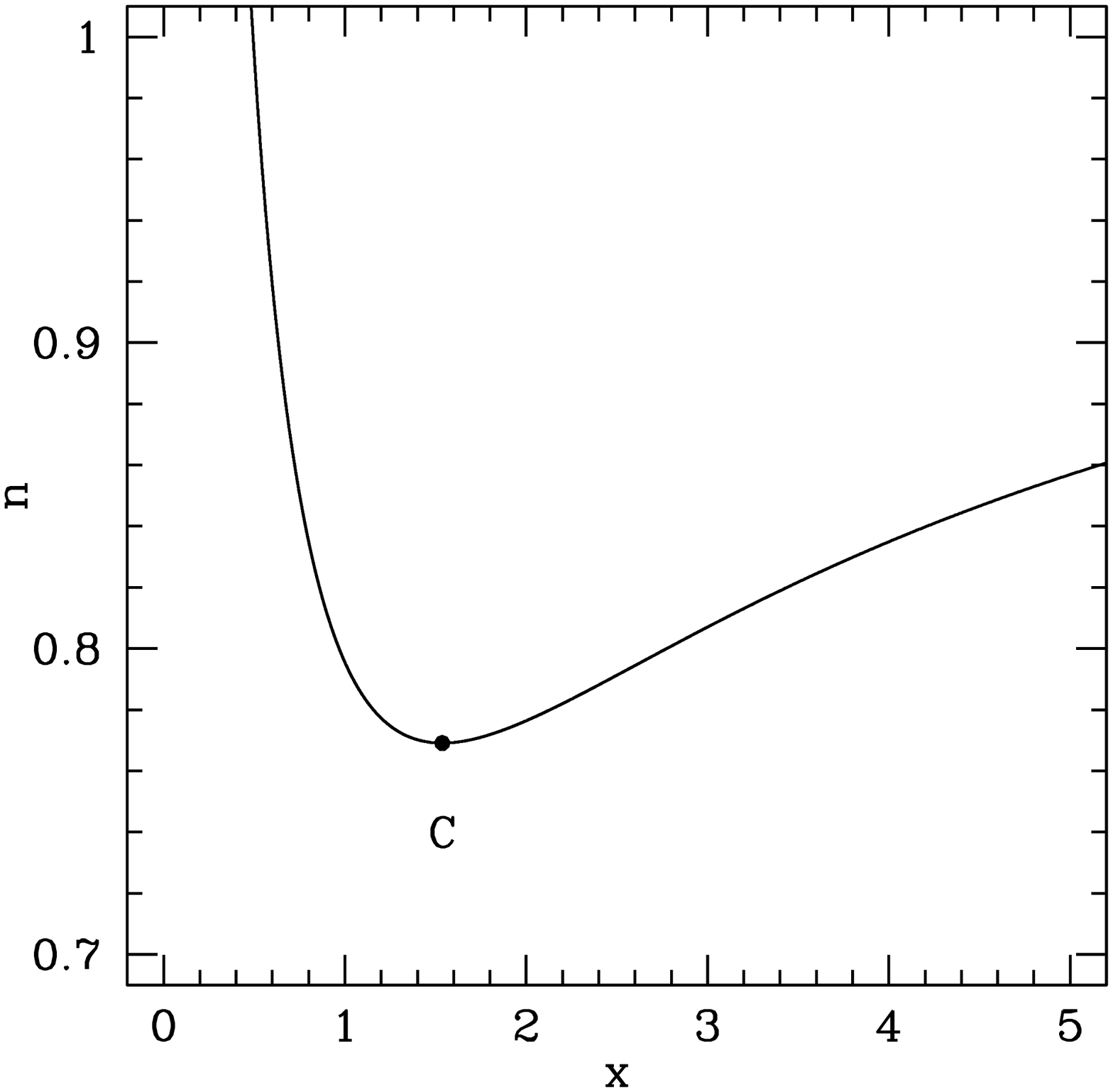}
\caption{
Nozzle function for the O~model ($s=0$).
When $s=0$ the critical point is determined by the position of the
minimum of the nozzle function.
``C'' denotes the critical point.
}
\label{fig_O_n}
\end{figure}

\begin{figure}
\plotone{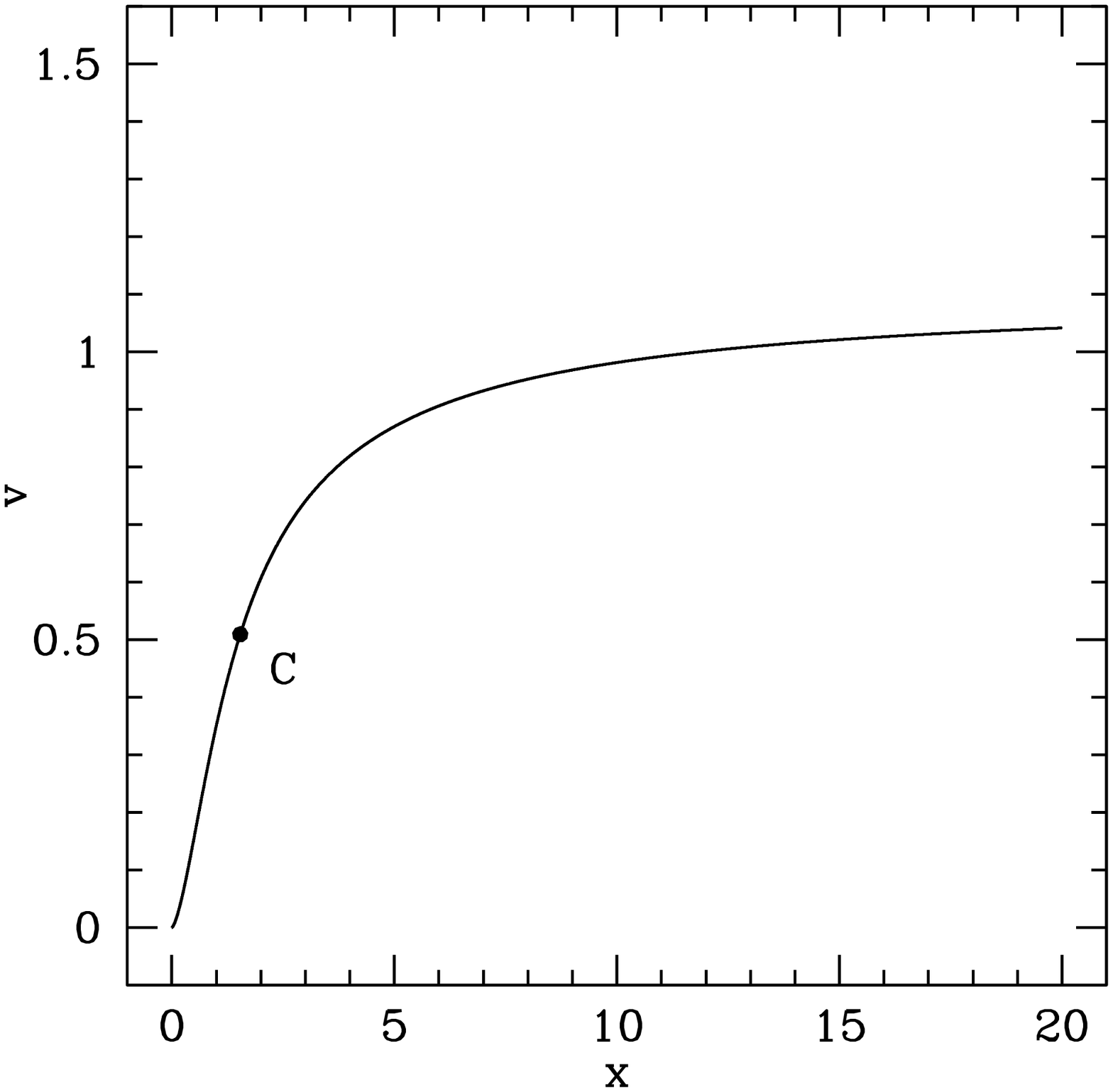}
\caption{
Velocity law for the O~model ($s=0$) obtained through
numerical integration of the equation of motion.
``C'' denotes the critical point.
}
\label{fig_O_v}
\end{figure}

\begin{figure}
\plottwo{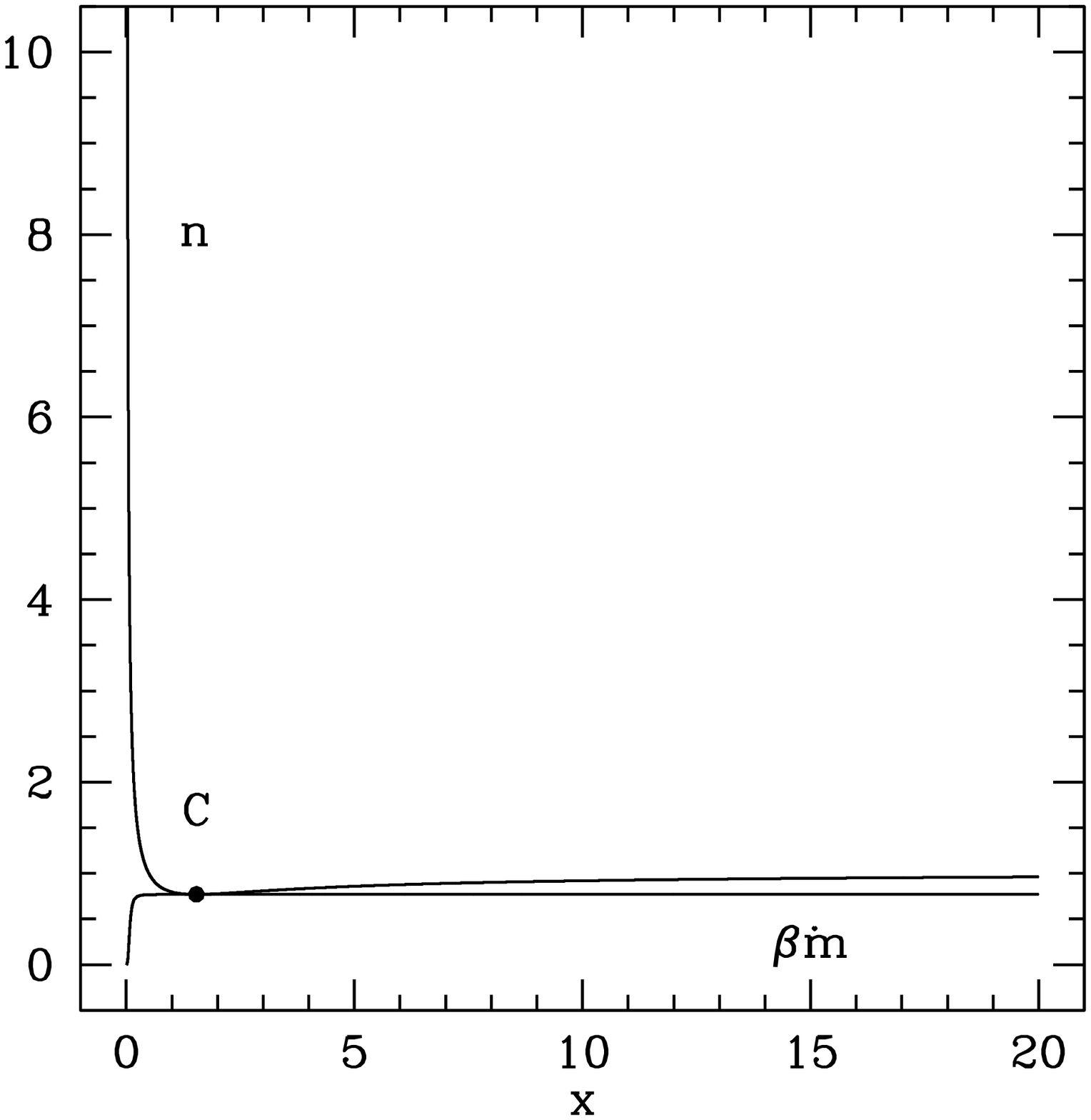}{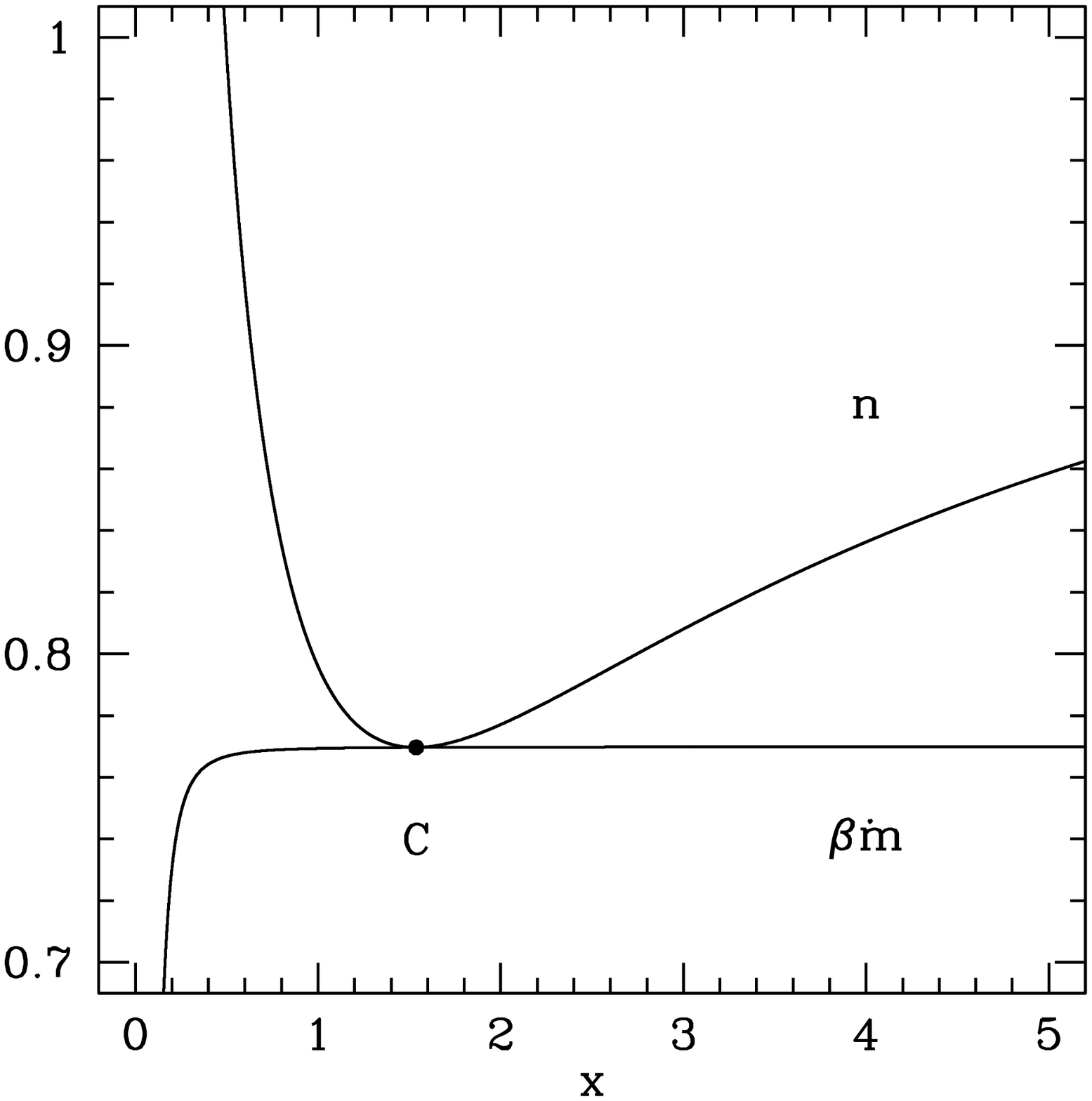}
\caption{
Nozzle function $n(x)$ and the product
$\beta(\omega[x]) \dot{m}$
for the O~model ($s=10^{-4}$).
Note that when gas pressure effects are included ($s>0$),
``$\beta(\omega[x])$'' is a monotonically increasing function
(equation~[\ref{equ_b}]);
thus (since $\dot{m}$ is a constant) it follows that the critical
point is shifted slight to the right of the nozzle minimum
(rather than {\it exactly} at the nozzle minimum,
 as when $s=0$).
``C'' denotes the critical point.
}
\label{fig_O_b}
\end{figure}

\begin{figure}
\epsscale{1.0}
\plotone{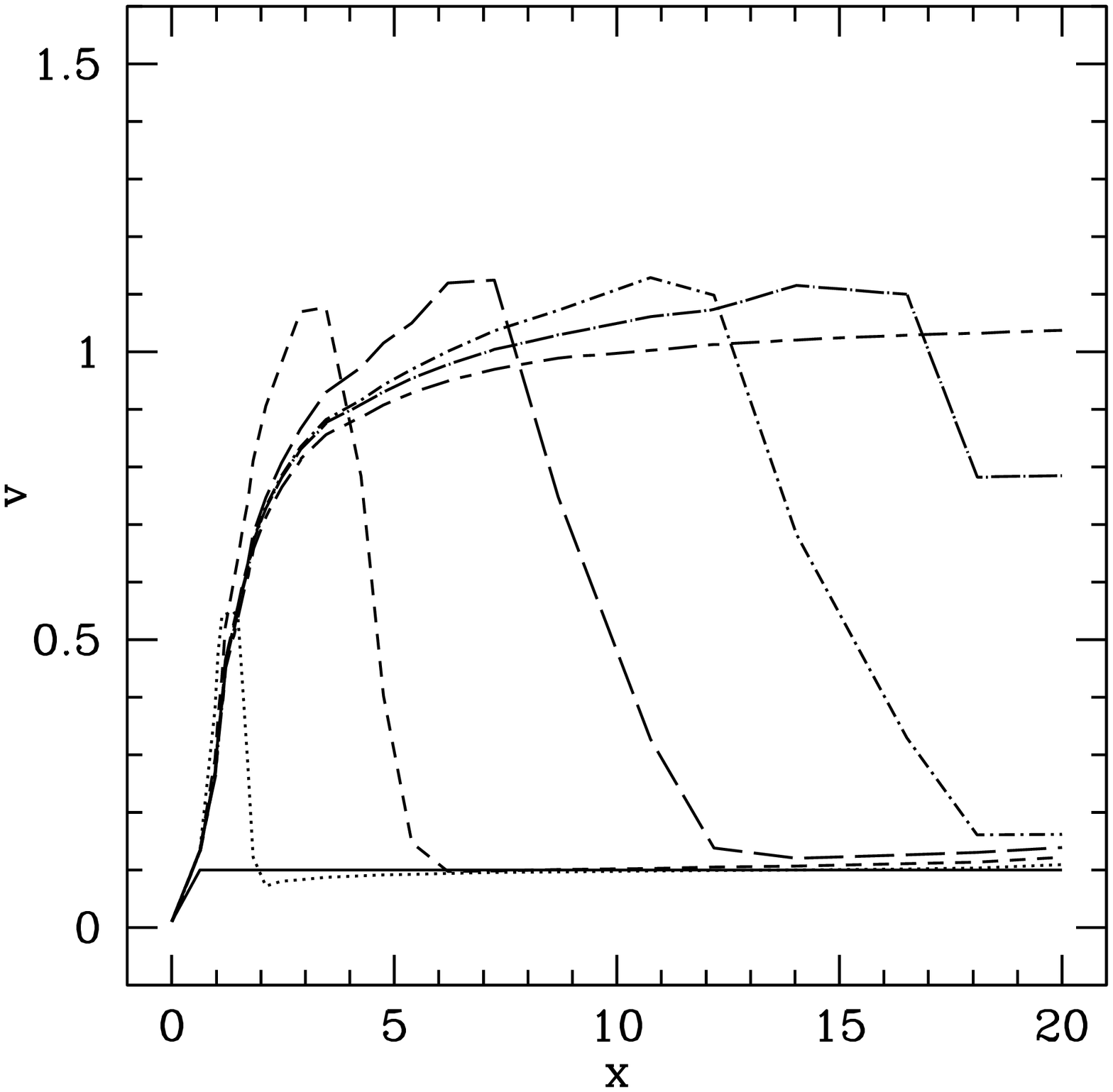}
\caption{
Time dependent velocity distribution for the O model
($s=10^{-4}$).
This plots shows that the steady solutions obtained through
the stationary numerical codes are stable.
In chronological order the plots are:
solid line,
dotted line,
short dashed line,
long dashed line,
short dot-dashed line,
long dot-dashed line,
and short dash-long dashed line.
The code arrives at a steady state after a small multiple of the
``crossing time''
(the time it would take a particle in the wind to travel from the
 the disk surface [$x=0$] to the right end of the spatial grid
 [$x=20$] once the steady state is achieved).
}
\label{fig_O_num}
\end{figure}

In Figure~\ref{fig_O_n} we present the nozzle function
$n(x)$
of the O~model
(with $s=0$).
The critical point is determined by the minimum of $n(x)$,
which in turn gives the value of the normalized wind mass loss rate
$\dot{m}$.
In Figure~\ref{fig_O_v} we present the velocity law of the O~model
derived through numerical integration starting from the critical point.
As with the I~model,
we compute the nozzle function and the velocity law of the O~model
for
$s=10^{-4}$,
but the corresponding corrections due to gas pressure were
insignificant,
so again we do not plot this to avoid redundancy.
Figure~\ref{fig_O_b} shows the nozzle function
$n(x)$
plotted simultaneously with the product
$\beta([\omega(x)]) \dot{m}$.
These two graphs intersect at,
and only at,
the critical point.
Since
$\beta(\omega[x])$
is a monotonically increasing function,
the critical point must be (slightly) to right of the nozzle minimum.
Figure~\ref{fig_O_num} shows the results from the time-dependent
simulation for the O~model.
Again the solid line is the initial velocity distribution at
$t=0$
and the short dash-long dashed line is the velocity distribution found
once the code arrives at a steady state.
Again this is in excellent agreement with the steady state solution
found through the stationary codes,
indicating that the steady solutions found through the stationary codes
for model~O are stable.

\section{Summary and Conclusions}

The objective of this work is to determine, in a manner independent
of from results of previous numerically-intensive 2.5D hydrodynamic
simulations,
whether steady line-driven disk wind solutions exist or not.
The motive behind this objective is to,
in turn,
determine whether line-driven disk winds could potentially account for
the wide/broad UV resonance absorption lines seen in CVs and QSOs.
In both types of objects,
it is observationally inferred that the associated absorption troughs
have steady velocity structure.

Our main conclusion is that if the accretion disk is steady,
then the  corresponding line-driven disk winds emanating from it can
also be steady.
We have confirmed this conclusion with more realistic
(and mathematically more elaborate)
models that we will present in a subsequent paper which implements the
exact flux distribution above a standard Shakura-Sunyaev disk.
This paper,
in particular,
has sought to emphasize on the underlying physics behind the steady
nature of line-driven disk winds through mathematically simple
models that mimic the disk environment.

The local disk wind mass loss rates and local disk wind velocity
laws represented by the standard simple models of this work are a
consequence of balancing the gravitational forces and the radiation
pressure forces.
Although gas pressure is present,
we find that in our models inclusion of gas pressure gives rise to only
minor corrections to the overall results.

The balance between gravitational and radiation pressure forces is
represented quantitatively through a nozzle function.
The spatial dependence of the nozzle function is a key issue in
determining the steady/unsteady nature of supersonic wind solutions.
In the case of a steady solution which neglects gas pressure effects,
the minimum of the nozzle function determines the corresponding wind
mass loss rate and the position of the minimum determines the critical
point.

In the vicinity of the disk,
gas pressure effects only produce minor corrections to the nozzle
function,
namely the critical point is shifted slightly to the right of the
nozzle minimum,
where the nozzle function has a positive derivative.
However,
in cases where the nozzle function is monotonically decreasing
(as in the S~model),
these minor corrections generate a minimum in the nozzle function at a
finite distance from the disk surface,
i.e.,
shifting the critical point from infinity to a finite distance.

The steady nature of line-driven disk winds found in this paper is
consistent with the steady nature of the streamline disk wind models of
Murray and collaborators
\citep{mur95,mur96,chi96,mur98},
and it is also consistent with the 2.5D time-dependent models of
Pereyra and collaborators
\citep{per97a,per00,hil02,per03}.

\acknowledgements

We wish to thank Kenneth G. Gayley and Norman W. Murray for many useful
discussions.

\appendix

\section{Nozzle Function and Critical Point for
         $s > 0$ and $0 < \alpha < 1$}
\label{sec_sg0}

For this case equation~(\ref{equ_motion_w}) is the equation of motion.
From an analysis similar to that presented by \citet{cas75},
this equation,
in the supersonic wind region,
has two solution branches which exist under the condition that

\begin{equation}
\label{equ_mdotn}
  \left(1 - {s \over w} \right) \dot{m}
\leq
  \alpha (1-\alpha)^{1 - \alpha \over \alpha}
    {(fa)^{1 \over \alpha} \over
     (ga-2s{da\over dx})^{1-\alpha \over \alpha}}
\;\;\;\; .
\end{equation}

\noindent
The two branches intersect at the critical point when both sides of
equation~(\ref{equ_mdotn}) are equal.
As is apparent from equation~(\ref{equ_mdotn}),
the critical point conditions now depend on
$x$
and
$\omega$
rather than only
$x$
as in cases where
$s=0$
(see \S\ref{sec_s0_a05} and \S\ref{sec_s0}).

The definition of the nozzle function
[cf. equation~(\ref{equ_n_s0})]
is now extended to include gas pressure in an isothermal wind using the
following equation

\begin{equation}
\label{equ_n}
  n(x)
\equiv
  \alpha (1-\alpha)^{1 - \alpha \over \alpha}
    {(fa)^{1 \over \alpha} \over
     (ga -2s{da\over dx})^{1-\alpha \over \alpha}} \; .
\end{equation}

\noindent
We also define the function
$\beta(\omega)$
as

\begin{equation}
\label{equ_b}
  \beta(\omega)
\equiv
  1 - {s \over w}
\;\;\;\; .
\end{equation}

\noindent
The reason for choosing equation~(\ref{equ_n}) as the definition for
the nozzle function when gas pressure effects are included are
threefold.
First,
the definition of
$\beta(\omega)$
[equation~(\ref{equ_b})]
allows for a simple relationship with the wind mass loss rate,
conserving its physical interpretation respect to the
$s=0$
case.
As we discuss below,
adding gas pressure effects will shift the critical point slightly to
the right of the minimum where the nozzle function has a positive
derivative.
Second,
equation~(\ref{equ_n}) reduces to the exact expression for the nozzle
function for the
$s=0$
case
(equation~[\ref{equ_n_s0}]).
Third,
equation~(\ref{equ_n}) depends only on
$x$
(not on $w$).
The importance of the
$x$-only-dependence
is that it allows for the development of mathematical/numerical
analysis of the problem without having to integrate the equation of
motion.
In particular,
one could determine through the above nozzle function whether or
not steady local wind solutions about a given spatial point exist,
without having to integrate the equation of motion.
This may lead to interesting future physical results,
as well as serve as a testing tool for numerical models aimed at
representing line-driven disk winds.

The condition for the existence of two solution branches can now be
written as

\begin{equation}
  \beta(\omega) \dot{m}
\leq
  n(x)
\;\;\;\; .
\end{equation}

\noindent
If the condition that
$\omega_c \gg s$
holds,
then we find that

\begin{equation}
  \dot{m}
\approx
  n(x_c)
\;\;\;\; .
\end{equation}

\noindent
Since we require a critical point type solution
(see \S\ref{sec_critical}),
a necessary condition for the critical point position is

\begin{equation}
\label{equ_nozzlecondition}
  \left. {dn \over dx} \right|_{x_c}
>
  0
\;\;\;\; .
\end{equation}

\noindent
We will demonstrate this in more detail in a future paper.
We emphasize that this applies when gas pressure effects are
included in an isothermal wind.
The actual critical point position is determined by a numerical
iterative process that successively integrates the equation of motion
until the condition
$\omega(x_s)=s$
is met.
This is equivalent to the iterative process used by
\citet{cas75}
in their original line-driven stellar wind paper.
A consequence of equation~(\ref{equ_nozzlecondition}) is that when gas
pressure effects are included,
the critical point is no longer
{\it exactly}
at nozzle minimum,
but in all the models we have explored,
the critical point position is shifted slightly to the right of the
nozzle minimum where the nozzle has a positive derivative.

\section{Necessity of a Critical Point for Steady Wind Solutions}
\label{sec_critical}

As discussed above,
the equation of motion [equation~(\ref{equ_motion_w})]
is integrable if upon integration one can always determine
(i.e., a steady solution exists;
       assuming the boundary condition can be met),

\begin{equation}
{d\omega \over dx}
=
{d\omega \over dx}(x,\omega)
\;\;\;\; .
\end{equation}

\noindent
Assuming the boundary condition can be met,
this means that a steady solution exists.

Viewing
$dw/dx$
as a function of variables
$x$
and
$\omega$
which satisfies equation~(\ref{equ_motion_w}),
one can divide the $x$-$w$ plane into five regions depending on whether
a solution for
$d\omega/dx$
exists.
We do this below in a manner equivalent to,
and following the notation of,
\citet{cas75}
for early type stars:

\begin{eqnarray}
\label{equ_region_map}
& & \hbox{Region~I:} \; \omega < s \;\; {\rm and} \;\;
     -ga + 2s{da\over dx} < 0 \; \hbox{: one solution};
    \nonumber
\\
& & \nonumber
\\
& & \hbox{Region~II:} \; \omega > s \;\; {\rm and} \;\;
    -ga + 2s{da\over dx} < 0 \;\; {\rm and} \;\;
    \beta(\omega)\dot{m} < n(x) \;
    \hbox{: two  solutions};
    \nonumber
\\
& & \nonumber
\\
& & \hbox{Region~III:} \; \omega > s \;\; {\rm and} \;\;
    -ga + 2s{da\over dx} > 0 \; \hbox{: one solution};
\\
& & \nonumber
\\
& & \hbox{Region~IV:} \; \omega > s \;\; {\rm and} \;\;
    -ga + 2s{da\over dx} < 0 \;\; {\rm and} \;\;
    \beta(\omega)\dot{m} > n(x) \; \hbox{: no solution};
    \nonumber
\\
& & \nonumber
\\
& & \hbox{Region~V:} \; \omega < s \;\; {\rm and} \;\;
    -ga + 2s{da\over dx} > 0 \;\; \hbox{: no  solution}.
    \nonumber
\end{eqnarray}

We now make the following five assumptions with respect to the
solution
$\omega(x)$:

 (1) $\omega(x)$
     increases monotonically
     (i.e., $d\omega/dx > 0$ [$dv/dx > 0$] throughout the wind);

 (2) the wind starts subsonic
     (i.e., the wind in the leftmost boundary is subsonic);

 (3) the wind ends supersonic
     (i.e., the wind in the rightmost boundary is supersonic);

 (4) the wind extends toward infinity
     (i.e., the rightmost boundary is infinity); and

 (5) $d\omega / dx$ is continuous
     (i.e., continuous velocity gradients).

\noindent
A wind solution must therefore start in Region~I,
since this is the only subsonic region which determines a value for
$d\omega/dx$.
Since we are assuming that the wind must end supersonic,
the solution must continue on to Region~II.
Region~II has two values for
$d\omega/dx$,
the lower/higher value corresponding to the lower/higher branch.
The solution in Region~I is continuous with the lower branch of
Region~II,
therefore the needed solution must go from Region~I to the lower branch
of Region~II as it goes from subsonic to supersonic.

Since the solution must extend toward infinity,
and assuming that asymptotically

\begin{equation}
x \to \infty : \cases{ a(x) \to x^2 \cr
                       g(x) a(x) \to {\rm const} }
\;\;\;\; ,
\end{equation}

\noindent
it follows that

\begin{equation}
x \to \infty : -ga + 2s{da\over dx} \to 4 x s
\;\;\;\; .
\end{equation}

\noindent
Thus,
toward infinity $-ga + 2s(da/dx) > 0$.
This implies that the solution must end in Region~III.

In turn,
the solution of Region~III is continuous with the upper branch of
Region~II.
Therefore,
at some large
$x$,
the solution must go from the upper branch of Region~II into
Region~III.
Continuity of
$d\omega/dx$
implies that the lower and higher branch solutions must intersect at a
point which is on the boundary between Region~II and Region~IV.
This intersection point is the critical point of the solution.

For a wind that starts subsonic,
reaches supersonic speeds,
and extends to infinity,
a solution to the 1D equations must then have the following sequence in
$x-\omega$
plane as
$x$
increases:

\hbox{ \hskip -24pt
\vbox{
\begin{eqnarray}
\label{equ_region_sol}
& & \hbox{Region~I: subsonic,} \; -ga + 2s{da \over dx} < 0;
    \nonumber
\\
& & \nonumber \\
& & \hbox{Region~II: supersonic,} \; -ga + 2s{da \over dx} < 0 \;
    \hbox{, lower branch,}
    \; \beta(\omega)\dot{m} < n(x);
    \\
& & \nonumber \\
& & \hbox{Region~II/Region~IV boundary: supersonic,} \;
    -ga + 2s{da \over dx} < 0, \; \hbox{\it critical point,}
    \; \beta(\omega)\dot{m} = n(x);
    \nonumber\\
& & \nonumber \\
& & \hbox{Region~II: supersonic,} \; -ga + 2s{da \over dx} < 0,  \;
    \hbox{upper branch,} \; \beta(\omega)\dot{m} < n(x);
    \nonumber \\
& & \nonumber \\
& & \hbox{Region~III: supersonic,} \; -ga + 2s{da \over dx} > 0 \ .
    \nonumber
\end{eqnarray}
}}

\noindent
We refer to solutions to the equation of motion with the above
characteristics as critical point type solutions.
We have therefore shown that a steady monotonically increasing
continuous wind solution that starts subsonic,
reaches supersonic speeds,
and extends to infinity must be a critical point type solution of the
equation of motion.


\clearpage

\clearpage

\end{document}